\documentclass[10pt,journal,a4paper]{IEEEtran}
\usepackage{amssymb,amsmath}
\usepackage{algorithm}
\usepackage{scalerel}
\usepackage{dsfont}
\usepackage[table]{xcolor}
\usepackage{mathbbol}
\usepackage{bbm}
\usepackage{algpseudocode}
\usepackage{cite}
\usepackage{graphicx,subfigure}
\usepackage{multirow}
\usepackage{psfrag}
\usepackage{url}
\usepackage{xcolor}
\usepackage[absolute,overlay]{textpos}
\usepackage{algpseudocode}
\usepackage{float}
\usepackage{lettrine}
\usepackage{indentfirst}
\usepackage{amsthm}

\usepackage{array}
\usepackage{psfrag}
\usepackage{url}
\usepackage{caption}
\usepackage[font=footnotesize]{caption}

\DeclareMathOperator*{\argminA}{arg\,min} % Jan Hlavacek
\makeatletter
\def\thickhline{%
  \noalign{\ifnum0=`}\fi\hrule \@height \thickarrayrulewidth \futurelet
   \reserved@a\@xthickhline}
\def\@xthickhline{\ifx\reserved@a\thickhline
               \vskip\doublerulesep
               \vskip-\thickarrayrulewidth
             \fi
      \ifnum0=`{\fi}}
\makeatother

\newlength{\thickarrayrulewidth}
\begin{document}
	
	\title{Robust Downlink Multi-Antenna Beamforming\\ with Heterogenous CSI:\\
	Enabling eMBB and URLLC Coexistence
}
	\author{
        \IEEEauthorblockN{Dian Echevarría Pérez, \IEEEmembership{Student Member, IEEE} \IEEEmembership{} %
        Onel L. Alcaraz López, \IEEEmembership{Member, IEEE},\\ Hirley Alves, \IEEEmembership{Member, IEEE}
        } 
        
		\thanks{The authors are with the Centre for Wireless Communications (CWC), University of Oulu, Finland. \{dian.echevarriaperez,  onel.alcarazlopez, hirley.alves\}@oulu.fi}

        \thanks{This research has been financially supported by Academy of Finland, 6G Flagship programme  (Grant no. 346208), HEXA-X (Grant Agreement no. 101015956), and the Finnish Foundation for Technology Promotion.}
    }  
    \maketitle
	
	\begin{abstract}
	 Two of the main problems to achieve ultra-reliable low-latency communications (URLLC) are related to instantaneous channel state information (I-CSI) acquisition and the coexistence with other service modes such as enhanced mobile broadband (eMBB). The former comes from the non-negligible time required for accurate I-CSI acquisition, while the latter, from the heterogeneous and conflicting  requirements of different nodes sharing the same network resources. In this paper, we leverage the I-CSI of multiple eMBB links and the channel measurement's history of a URLLC user for multi-antenna beamforming design. Specifically, we propose a precoding design that minimizes the transmit power of a base station (BS) providing eMBB and URLLC services with signal-to-interference-plus-noise ratio (SINR) and outage constraints, respectively, by modifying existing I-CSI-based precoding schemes to account for URLLC channel history information. Moreover, we illustrate and validate the proposed method by adopting  zero-forcing (ZF) and the transmit power minimization (TPM) precoding with SINR constraints. We show that the ZF implementation outperforms TPM in adverse channel conditions as in Rayleigh fading, while the situation is rapidly reversed as the channel experiences some line-of-sight (LOS). Finally, we determine the confidence levels at which the target outage probabilities are reached. For instance, we show that outage probabilities below $10^{-3}$ are achievable with more than 99$\%$ confidence for both precoding schemes under favorable LOS conditions with 16 transmit antennas and 500 samples of URLLC channel history.
	\end{abstract}
\section{Introduction}
Ultra-reliable low-latency communication (URLLC) is a key operation mode in current and future wireless communication networks. Many envisioned applications require reliability levels close to those offered by wired networks, \textit{i.e.}, error probabilities below $10^{-5}$ almost 100 $\%$ of the time, and latency levels below 10 ms \cite{popovski2019wireless}. For instance, automotive communications require a user plane reliability of $10^{-5}$ and end-to-end (E2E) latencies below 5, 10, and 20 ms for assisted, cooperative, and tele-operated driving, respectively. Motion control in industrial processes requires also reliability levels of $10^{-5}$ and E2E latencies up to 1 ms \cite{lorca2017deliverable}. 

In practice, it is difficult/challenging to meet both reliability and latency requirements simultaneously. Communication latency can be reduced by shortening the transmission time interval, using non-slot or mini-slot scheduling policies, and/or employing uplink (UL) grant-free transmissions \cite{li20185g}.
On the other hand, diversity techniques, \textit{e.g.}, time, frequency, or spatial diversity, are key to achieving high reliability levels. Notice that message retransmission may be seen as a time diversity technique to improve reliability, but at the cost of higher latencies \cite{elayoubi2019radio}, while frequency diversity may not always be available given spectrum-sharing/slicing constraints. Instead, spatial diversity, whose availability is increasing due to the rise of multiple-input multiple-output (MIMO) communications and node densification, is  often the most appealing to support ultra-reliable services \cite{popovski2019wireless}. Specifically, precoding/combining allows increasing the signal power and/or suppressing the interference from other users or base stations (BS).  This, in turn, improves the signal-to-interference-plus-noise ratio (SINR) statistics, the data decoding performance, and consequently leads to reliability enhancements. 
For both, precoding and combining procedures, channel state information (CSI) is commonly needed. However, instantaneous CSI (I-CSI) of URLLC links might be too costly to acquire if the latency constraints are too tight \cite{9269936}. The energy consumption is also a critical aspect for the CSI acquisition. Low-energy devices may not afford to participate frequently in CSI acquisition procedures due to the corresponding non-negligible energy expenditure. Hence, other approaches must be considered, e.g.,  exploiting channel statistics (\textit{e.g.}, mean and covariance matrix) instead of instantaneous channel realizations. These statistics do not change regularly, especially in slow fading scenarios where the coherence time is larger than the delay requirements of the application. In general, the use of channel statistics may be a suitable option when delay and/or energy constraints are strict, while I-CSI acquisition procedures may be carried out when the conditions are more favorable. 

 Several works have been conducted exploiting the channel statistics for multi-antenna precoding design. For instance, the authors in \cite{bornhorst2012beamforming} considered the problem of transmit power minimization with channel covariance-based beamforming in  multicast scenarios. The work in \cite{tabikh2017beamforming} focused on the beamforming design for weighted sum-rate maximization using combined channel mean and covariance information. The authors in \cite{qiu2017downlink}  addressed the problem of downlink (DL) precoding design with mixed  statistical and imperfect I-CSI in massive MIMO systems. They proposed extended zero-forcing (ZF) and maximal ratio transmission (MRT) methods to minimize the total transmit power. However, none of these works focused on supporting URLLC services. In this regard, the work in \cite{bana2018ultra} exploited the sparsity of the propagation channel and relied on the estimation of a small number of channel coefficients for the beamforming design. Meanwhile, authors in \cite{nasir2020resource} proposed a beamformer that maximizes the users' minimum rate in an interference limited multi-user system with short packet transmissions and URLLC constraints. 
 
 Notice that URLLC services will necessarily coexist with other operation modes such as enhanced mobile broadband (eMMB). Such coexistence is of paramount importance in current and future wireless communication networks and has attracted a lot of attention in recent years \cite{pedersen2017punctured,wang2021joint,anand2020joint,alsenwi2019embb, yin2020multiplexing,bairagi2020coexistence,alsenwi2021intelligent}. On the one hand, URLLC users transmit usually at low data rates, are characterized by intermittent activation patterns mainly associated with external events, such as alarms, and require transmitting short messages, in the order of a few tens of bytes. On the other hand, eMBB users require high data rates with steady activation patterns \cite{popovski20185g}. These fundamental differences make the network design to meet all the requirements, a challenging task. Therefore, to cope with the quality-of-service (QoS) requirements of all nodes, it is necessary to develop efficient multiplexing techniques for URRLC and eMBB users. DL channel preemptive scheduling, where the data of the URLLC user is transmitted immediately and  overwrites the current transmission intended for the eMBB user, is an example technique. The advantage of this method is that the transmission intended for the URLLC user does not have to wait for the scheduled slot, and the drawback is a performance degradation of the eMBB user \cite{pedersen2017punctured}. To tackle this issue, the authors in \cite{anand2020joint} were the first to explore resource allocation for joint scheduling of URLLC and eMBB traffic using puncturing/superposition based methods. Therein, they investigated various models, i.e., linear, convex, and threshold-based, for describing the impact of the URLLC traffic load on the rate loss of the eMBB users. In \cite{bairagi2020coexistence}, the authors focused on the co-scheduling  of URLLC and eMBB traffic based on puncturing, and aimed to maximize the minimum expected achieved rate of eMBB users while fulfilling the URLLC traffic demands. Meanwhile, the multiplexing
of eMBB and URLLC traffic in the DL channel was analyzed in \cite{yin2020multiplexing}. Specifically, a resource allocation problem in each mini-slot was formulated as an integer programming problem to maximize an eMBB utility function  while satisfying URLLC constraints. The work in \cite{alsenwi2019embb} presented a risk-sensitive based formulation that allocates relatively more URLLC traffic to the network resources reserved for eMBB users with higher data rates.  In \cite{alsenwi2021intelligent}, a resource slicing optimization problem was formulated to maximize the eMBB data rate while satisfying the performance requirements of the URLLC traffic. Therein, a deep reinforcement learning framework, including eMBB resource allocation followed by URLLC scheduling, was proposed to solve the problem. 
\subsection{Motivation}

 There are still some open challenges for efficiently enabling URLLC-eMBB coexistence, which has not been considered in the previous works. For instance, how to efficiently multiplex the correspondingly heterogeneous services in time, frequency, and space. Even more challenging is how to design the spatial precoding if the latency constraint is too tight such that I-CSI cannot be acquired. The CSI history of URLLC links may be leveraged to address such an issue.  Interestingly, relying on a limited number of past channel measurements to design URLLC-supporting precoders, although appealing, has not been considered in the literature to the best of authors' knowledge. Still, exploiting channel history has already proven valuable to enable URLLC \cite{mahmood2020predictive,azari2019risk,huang2019machine,samarakoon2020predictive}. For instance, the authors in \cite{mahmood2020predictive} proposed an interference prediction algorithm for supporting URLLC. Specifically, the interference dynamics were modeled as a discrete-time
Markov chain with state transition probability matrices being estimated  using past interference measurements. Meanwhile, machine-learning (ML) mechanisms were proposed in \cite{azari2019risk,huang2019machine,samarakoon2020predictive} to support URLLC in different scenarios. In \cite{azari2019risk}, the authors studied the coexistence design challenges of scheduled and non-scheduled URLLC traffic, and presented a distributed risk-aware ML solution for the corresponding radio resource management problem (RRM). In \cite{huang2019machine}, the authors introduced ML and fountain codes into millimeter wave hybrid access, and proposed an adaptive channel assignment method for URLLC. The work in \cite{samarakoon2020predictive} characterized the wireless connectivity over dynamic channels via statistical learning methods, and measured the reliability of wireless connectivity in terms of the probability of channel blocking events. However, notice that the main drawback of ML-based mechanisms in the context of URLLC lies in the big data requirements, e.g., for model training, specially in dynamic environments,
and/or the exploitation of latency-unfriendly feedback/signaling channels. 

Furthermore, I-CSI might not be available for URLLC under tight latency constraints. Thus, herein we focus on exploiting URLLC channel history for precoding design in heterogeneous scenarios. Notice that the system may provide service to a more significant number of devices by enabling a harmonious coexistence of URLLC and eMBB services in the same resource blocks (time-frequency) \cite{wang2021joint}.
  
\subsection{Contributions}
In our work, we focus on transmit power minimization through precoding design in heterogeneous scenarios with coexisting URLLC and eMBB DL users, and no I-CSI availability for URLLC services. Specifically, our contributions are four-fold:
\begin{itemize}
    \item We formulate a precoding optimization problem concerning transmit power minimization while ensuring URRLC and eMMB coexistence with different CSI availability. Furthermore, we exploit the Chernoff bound to stochastically model, impose, and guarantee the reliability requirements of the URLLC user based on its channel history.
    
    \item We propose an algorithm that leverages existing I-CSI-based precoding methods to solve the optimization problem. This allows taking advantage of efficient state-of-art precoders with relatively low implementation difficulty. We show that the algorithm complexity grows with the number of iterations $\zeta$ and the third power of the total number of users $(K+1)^3$.
    
    \item We evaluate the performance of the proposed algorithm with ZF precoding and the transmit power minimization (TPM) precoding with per-user SINR constraints. We show that ZF outperforms TPM in poor channel conditions, e.g., Rayleigh fading, while TPM exhibits superior performance in more deterministic channels, e.g., Rician fading with a significant $\kappa$ factor.

     \item We analyze the impact of $\zeta$, the Chernoff bound auxiliary variable $r$, and the number of past channel measurements $L$ on the system performance. We show that large values of $r$ and $\zeta$ may reduce the transmit power significantly but at the cost of affecting the reliability performance in practice. We also determine the confidence levels required to reach the target outage probabilities.  For instance, outage probabilities below  $10^{-3}$ are achievable with more than 99\% confidence 
    for both precoding methods in favorable line-of-sight (LOS) conditions with 16 transmit antennas at the BS, and given $L=500$ samples of URLLC channel history.
    \end{itemize}
    
The work is structured as follows. In Section \ref{system_model}, we describe the system model, main assumptions, and formulate the optimization problem. In Section \ref{section_3}, we present the history-based beamforming design, the proposed algorithm, and discuss ZF and TPM-based implementations. In Section \ref{section_4}, we illustrate numerical results and validate the performance of the proposed algorithm. Finally, Section \ref{section_5} concludes the paper.

\begin{table}[t!]
    \centering
    \caption{Main symbols used throughout the paper}
    \label{table_0}
    \begin{tabular}{l  l}
        \hline
        \textbf{Symbol} & \textbf{Definition} \\
            \hline    
            $M$ & number of transmit antennas at the BS\\
            $K$ & total number of eMBB users\\
            $\mathbf{h}_k$ & channel vector between the BS and user $k$\\
            $\tilde{\mathbf{h}}_{0,l}$ & $l-$th past channel measurement of the URLLC user\\
            $\mathbf{w}_k$ & precoder intended to user $k$\\
            $\mathbf{u}_k$ & normalized precoder intended to user $k$\\
            $\gamma_{k}$ & SINR at user $k$ \\
            $\gamma_{k}^{tar}$ & SINR target at user $k$\\
            $\sigma^2$ & noise power\\
            $s_k$ & complex baseband signal corresponding to user $k$\\
           $p_k$ & power allocated to user $k$\\
           $p_{max}$ & maximum transmit power at the BS\\
          $\xi$ & outage probability target\\
          $L$ & number of past measurements of the URRLC channel\\
          $\hat{\mu}$ & sample mean in the Chernoff bound framework\\
          $\hat{s}$ & sample standard deviation in the Chernoff bound framework\\
          $\mu_{UB}$ & upper bound of the population mean\\
          $\alpha$ & confidence of the upper bound $\mu_{UB}$\\
          $\zeta$ & number of generated channel coefficients\\
          $d_r$ & radius of the network deployment area\\
        $\delta$ & path loss exponent\\
          $\psi$& path gain\\
          $\mathcal{O}_u$& outage probability of the URLLC user\\
          $CV$ & confidence value of $\log_{10}\mathcal{O}_u$\\
          $MV$ & estimated mean value of $\log_{10}\mathcal{O}_u$\\
          $SD$ & estimated standard deviation of $\log_{10}\mathcal{O}_u$\\
            \hline
    \end{tabular}
\end{table}

    \textbf{Notation} Uppercase and lowercase boldface letters denote matrices and vectors, respectively. Superscript $(\cdot)^*$ denotes complex conjugate, $(\cdot)^H$ depicts the Hermitian operator, and $(\cdot)^{-1}$ represents the matrix inverse operation. $||\cdot||$ represents the norm of a vector, and $||\cdot||_F$ the Frobenius norm. Moreover, $\mathcal{CN}(\mathbf{v},\mathbf{R})$ denotes a complex Gaussian distribution with mean vector $\mathbf{v}$ and covariance matrix $\mathbf{R}$. Finally, $\mathcal{U}(\upsilon_1,\upsilon_2)$ represents a uniform distribution in the range $[\upsilon_1,\upsilon_2]$, and $Q(\cdot)$ depicts the Q-function. 

\section{System model}\label{system_model}

\begin{figure}[t!]
    \centering
    \includegraphics[width = 0.9\columnwidth]{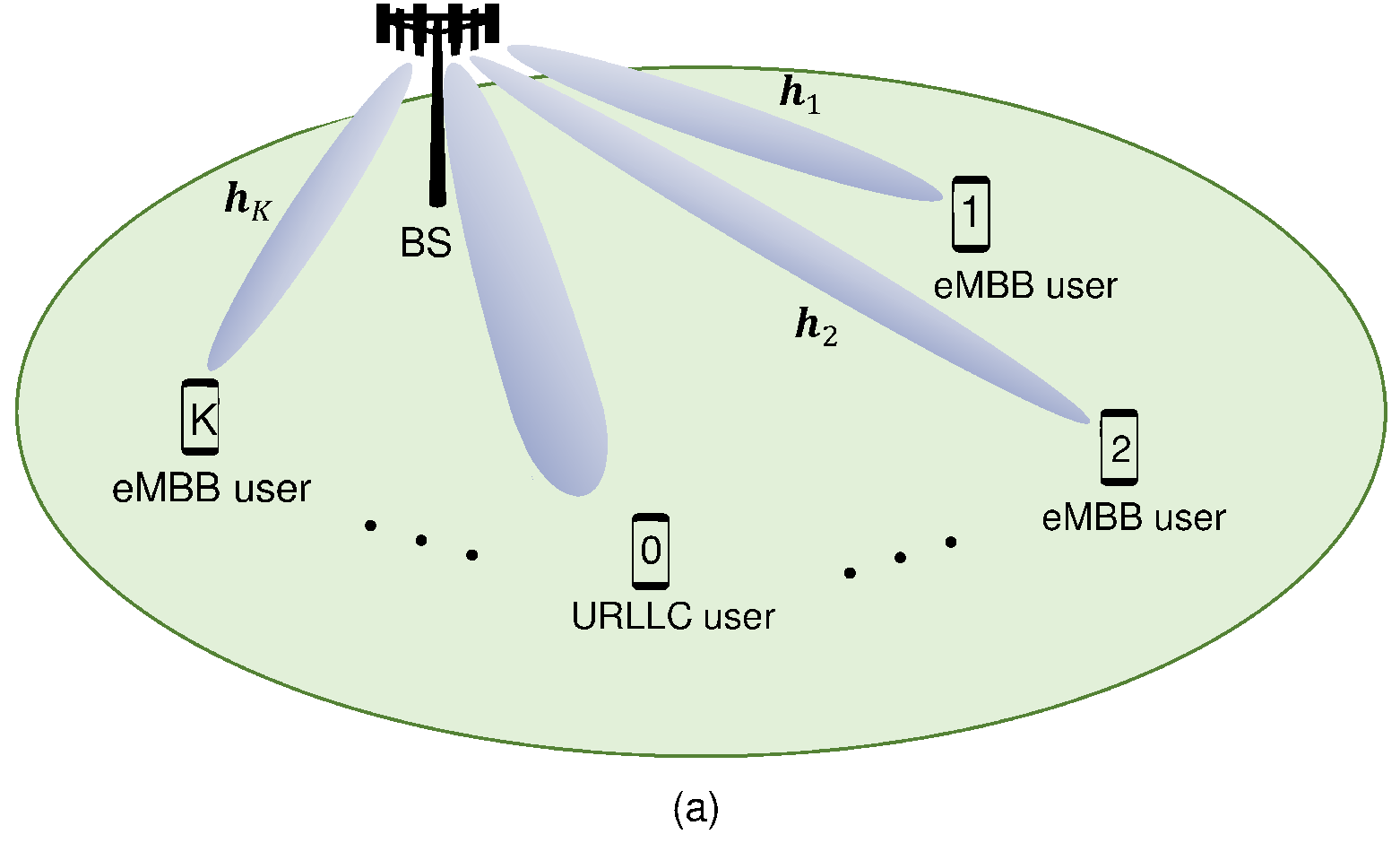}
    \includegraphics[width = 0.9\columnwidth]{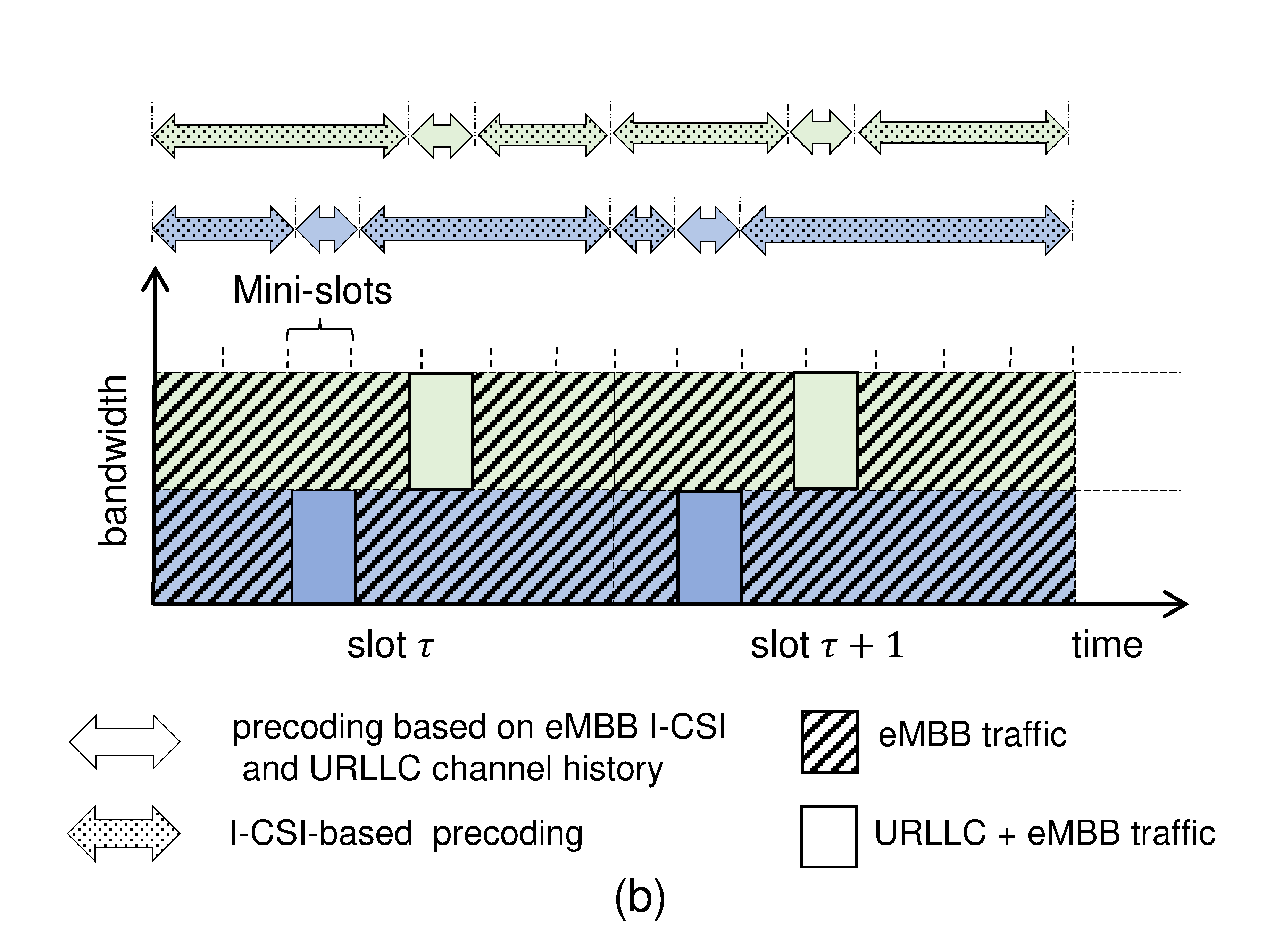}
    \caption{(a) System model, and (b) multiplexing scheme. A BS serves one URLLC node and a set of $K$ eMBB users within a time-frequency block in the DL channel. The transmit beams are perfectly focused on the direction of the eMBB users since they exploit the available I-CSI, but the beam towards the URRLC user is not perfectly oriented, neither sharp, since its design depends on imperfect channel statistics. Pure eMBB transmissions are served via I-CSI-based precoding, while the BS implements a hybrid precoding when a URLLC service is triggered and will coexist with the eMBB services, for which it leverages  eMBB I-CSI and URLLC CSI history.}
    \label{fig:System_model}
\end{figure}

We consider a scenario where a BS equipped with $M$ antennas spatially multiplexes one URLLC user and $K$ eMBB users, with $K+1\le M$, within a resource block, \textit{i.e.}, time-frequency resource, as depicted in Fig. \ref{fig:System_model} (a). Nevertheless, notice that the BS may serve multiple URLLC users within different resource blocks as shown in Fig. \ref{fig:System_model} (b), while herein, we focus on a single resource block  operation without loss of generality. Scheduled eMBB users are continuously receiving data in the DL, and their QoS is guaranteed provided that their target SINR, $\gamma_{k}^{tar}$ for user $k=1,2,...K$,  is surpassed. The I-CSI of the eMBB users is obtained via training/feedback methods prior to the DL transmissions, and is assumed to be always known at the BS. On the other hand, URLLC users with strict latency and reliability requirements do not receive data all the time in the DL, which agrees with the typically sporadic data transmissions in many URLLC use cases  \cite{popovski20185g}. Moreover, cooperation for I-CSI is unaffordable under the considered  latency constraints due to the delays that the required procedures would introduce \cite{9269936}. Therefore, I-CSI of the URLLC link is not available at the BS before a DL transmission takes place. Instead, we assume that if the URLLC user does not receive critical information, it participates in frequent CSI acquisition also via training/feedback, while the BS performs I-CSI-based transmit beamforming to serve the $K$ eMBB users. 
  On the other hand, when a transmission to a URLLC user is required, the BS uses both the I-CSI of the eMMB users and the past channel measurements of the corresponding URLLC channel for the  beamforming design.
 
  \subsection{Signal model}
 The received signal at the $k-$th user at a certain time-frequency resource is given by 
\begin{equation}
    y_k = \mathbf{h}_k^H\mathbf{w}_ks_k+\sum_{i\neq k}\mathbf{h}_k^H\mathbf{w}_is_i+n_k,
\end{equation}
where $\mathbf{h}_k\in \mathbb{C}^{M\times 1}$ represents the complex vector containing the channel coefficients between the $M$ antennas of the BS and user $k$, $\mathbf{w}_k\in \mathbb{C}^{M\times 1}$ is the precoding vector intended to user $k$, and $s_k$ denotes the complex baseband signal transmitted to user $k$ such that $\mathbb{E}\{s_k^*s_k\} = 1$ and $\mathbb{E}\{s_k^* s_i\}=0$    $\forall k\ne i$. Finally, $n_k\sim \mathcal{CN}(0, \sigma^2)$ represents the additive white Gaussian noise. We use indexing from 0 to $K$  with index 0 referring to the URLLC user. The SINR at user $k$ is
\begin{equation}\label{eq2}
   \gamma_k(\{{\mathbf{w}_k}\}) = \frac{|\mathbf{h}_k^H\mathbf{w}_k|^2}{\sum_{i\neq k}|\mathbf{h}_k^H\mathbf{w}_i|^2+\sigma^2}.
\end{equation}
\subsection{Problem formulation}
Herein, we aim at configuring the transmit precoding that satisfies the URLLC and eMBB related QoS constraints of instantaneous DL transmissions with minimum power. For this, we set the following optimization problem
\begin{subequations}\label{P1}
	\begin{alignat}{2}
	\mathbf{P1:}\ \ &\underset{\{\mathbf{w}_i\}_{\forall i}}{\mathrm{minimize}}  &\ \ &\
	\sum_{i=0}^K ||\mathbf{w}_i||_2^2\label{P1:a}\\
	&\text{subject to}   &      & \mathcal{O}_u=\text{P}\big\{\gamma_0(\{\mathbf{w}_k\})<\gamma_0^{tar}\big\} \leq \xi, \label{P1:b}\\
	&   &  & \gamma_k(\{\mathbf{w}_k\})>\gamma_{k}^{tar}, \ \ k = 1, 2,\hdots, K, \label{P1:c}
	\end{alignat}
\end{subequations}
where $\gamma_0(\{\mathbf{w}_k\})$ represents the  receive SINR at the URLLC user. Moreover, $\gamma_0^{tar}$
is the SINR threshold required to achieve a successful URLLC transmission, $\xi$ is maximum allowable outage probability, and $\gamma_{k}^{tar}$ depicts the required SINR for  proper operation of eMBB user $k$. In general, \eqref{P1:b} guarantees that the outage probability of the URLLC user, $\mathcal{O}_u$, is kept below the outage target. 

Note that the objective function \eqref{P1:a} in $\mathbf{P1}$ is convex. On the other hand, the constraint \eqref{P1:c} is usually rearranged to solve the problem (without \eqref{P1:b}) via second order cone programming or semi-definite relaxation, among others. However, constraint \eqref{P1:b} is not convex and difficult to handle in general since the channel distribution is assumed unknown, and therefore, an expression for the outage probability is not available, which is typical in practical systems. Even if we consider an empirical approximation to the channel fading probability distribution, a large number of measurement samples, \textit{i.e.}, at least $10/\xi$, would be required, and the mathematical complexity for solving the optimization problem would be high anyway. Since the I-CSI of the URLLC link is not available to solve $\mathbf{P1}$, we rely on past URLLC channel measurements $\{\mathbf{\tilde{h}}_{0,1} \ \mathbf{\tilde{h}}_{0,2}\hdots\mathbf{\tilde{h}}_{0,L}\}$, and the I-CSI of eMBB users to design all precoders.
\section{History-based bemforming design}\label{section_3}
The main difficulty in solving $\mathbf{P1}$ lies in efficiently addressing the constraint \eqref{P1:b} given a limited number $L$ of past channel samples of the URLLC link. To address this, herein, we apply the Chernoff bound to reformulate \eqref{P1:b} as
\begin{equation}\label{eq:4}
 \mu = e^{r\gamma_0^{tar}}\mathbb{E}\{e^{-r\gamma_{0}(\mathbf{\{w}_k\})}\}\ \leq \xi, \ \ \forall r>0,
\end{equation}
where the expectation $\mathbb{E}\{e^{-r\gamma_{0}(\{\mathbf{w}_k\})}\}$ is taken over $\mathbf{h}_0$, and $r$ is an auxiliary variable. Since a set of $L$ channel realizations is available, the expectation could be approximated to the sample mean as
\begin{equation}\label{eq5}
\hat{\mu} = \frac{1}{L}\sum_{j=1}^Le^{r(\gamma_0^{tar}-\gamma_{0,j}(\{\mathbf{w}_{k}\}))},
\end{equation}
where $\gamma_{0,j}$ is given by
\begin{align}
    \gamma_{0,j}(\{{\mathbf{w}_k}\}) = \frac{|\tilde{\mathbf{h}}_{0,j}^H\mathbf{w}_k|^2}{\sum_{i\neq k}|\tilde{\mathbf{h}}_{0,j}^H\mathbf{w}_i|^2+\sigma^2}.
\end{align}
Note that $\hat{\mu}\xrightarrow[]{}\mu$ holds only when $L\xrightarrow[]{}\infty$ due to the law of large numbers. Therefore, since the value of $\hat{\mu}$ in \eqref{eq5} might significantly deviate from the population mean for a limited number $L$ of channel measurements, we proceed as follows.

Observe that the  population mean can be bounded with $100\times\alpha\%$ confidence using percentiles over the distribution of the sample mean as follows
\begin{align}\label{eq6}
    \text{P}\{\mu_{UB}\geq\mu \}\!&= \alpha\nonumber\\
    \text{P}\bigg\{\!(\hat{\mu}\!-\!\mu_{UB})\frac{\sqrt{\!L}}{\hat{s}}\leq(\hat{\mu}\!-\!\mu)\frac{\sqrt{\!L}}{\hat{s}}\bigg\}\! &= \alpha\nonumber\\
    \text{P}\bigg\{(\hat{\mu}-\mu_{UB})\frac{\sqrt{L}}{\hat{s}} \leq \varphi\bigg\} \!&= \alpha\nonumber\\
    F_{\varphi}\bigg((\hat{\mu}-\mu_{UB})\frac{\sqrt{L}}{\hat{s}}\bigg)&= 1- \alpha\nonumber\\
     \mu_{UB}\!&\triangleq \hat{\mu}\!-\!\frac{\hat{s}}{\sqrt{L}}F_{\varphi}^{-1}(1\!-\!\alpha) ,
\end{align}
where $\mu_{UB}$ is an upper bound of the population mean, $\hat{s}$ is the sample standard deviation, $F_{\varphi}(\cdot)$ depicts the cumulative distribution function of $\varphi = (\hat{\mu}-\mu)\frac{\sqrt{L}}{\hat{s}}$, and $F_{\varphi}^{-1}(\cdot)$ its inverse. The sample standard deviation is given by
\begin{align}
\hat{s} = \sqrt{\frac{1}{L-1}\sum_{j=1}^L \big(e^{r(\gamma_{0}^{tar}-\gamma_{0,j}(\{\mathbf{w}_{k}\}))}-\hat{\mu}\big)^2}.
\end{align}
The sample mean $\hat{\mu}$ tends to follow a normal distribution around $\mu$ as $L$ increases. However, since the population standard deviation  is unknown we must rely on $\hat{s}$. In such case, the Student's $t$ distribution, which converges to a normal distribution when the number of samples goes to infinity, must be used. Notice that this distribution turns out to be more useful than the normal distribution for smaller number of samples due to its heavier tail.  Then, the problem $\mathbf{P1}$ is re-written as
\begin{subequations}\label{P2}
	\begin{alignat}{2}
	\mathbf{P2:}\ \ &\underset{\{\mathbf{w}_i\}\forall i , r}{\mathrm{minimize}}  &\ \ &\
	\sum_{i=0}^K ||\mathbf{w}_i||_2^2\label{P2:a}\\
	&\text{subject to}   &      & \mu_{UB}\leq\xi, \label{P2:b}\\
	&   &  & \gamma_k(\{\mathbf{w}_k\})>\gamma_{k}^{tar}, \ k= 1,2,\hdots,K, \label{P2:c}\\
	&   &  & r>0\label{P2:d}.
	\end{alignat}
\end{subequations}

Note that $\mathbf{P2}$ is still difficult to solve analytically, while common solvers (\textit{e.g.}, genetic algorithm (GA), particle swarm optimization (PSO)) do not often provide feasible solutions. This comes from the high non-linearity of \eqref{P2:b}, and the difficulty to configure appropriately the solvers, which often requires lots of computational resources. Alternatively, we propose taking advantage of existing solutions in the literature related to problem $\mathbf{P1}$ that do not consider the reliability constraint. Next, we provide specific details, which includes a heuristic for solving  $\mathbf{P2}$.
\subsection{Proposed algorithm} 
The proposed algorithm leverages existing I-CSI -based precoding schemes (such as those in \cite{bjornson2014optimal, khandaker2018signal}), but here, they are fed with random channel vectors. The random channel vectors are generated using statistics obtained from the channel history of the URLLC link, specifically, the sample mean $\bar{\mathbf{m}}$ and sample covariance matrix $\mathbf{C}$, which are given by
\begin{align}\label{sample_mean}
    \bar{\mathbf{m}} &= \frac{1}{L} \sum_{j=1}^L \tilde{\mathbf{h}}_{0,j},\\
    \mathbf{C} &= \frac{1}{L-1}\sum_{j=1}^{L}(\tilde{\mathbf{h}}_{0,j}-\bar{\mathbf{m}})(\tilde{\mathbf{h}}_{0,j}-\bar{\mathbf{m}})^H.
\end{align}
With these statistics, we generate $\zeta$ random channel vectors such that 
\begin{align}\label{eq10}
    \mathbf{h}_{0,t} \sim  \mathcal{CN}(\bar{\mathbf{m}}, \mathbf{C}), \ \text{for} \ t = 1,2,\hdots,\zeta.
\end{align}
   Then, for each generated vector, we define a channel matrix
\begin{align}\label{h_matrix}
    \mathbf{H}_t =[\mathbf{h}_{0,t} \ \mathbf{h}_{1}\hdots \mathbf{h}_{K}],
\end{align}
    where the last $K$  columns correspond to the I-CSI of the eMBB users. To compute the individual precoders $\mathbf{w}_{k,t}$, we  first determine separately the normalized precoders and their associated power $p_{k,t}$ such that  $\mathbf{w}_{k,t} = \sqrt{p_{k,t}}\mathbf{u}_{k,t}$.
    
    The $t-$th matrix of normalized precoders has the structure     \begin{align}\label{U_structure}
       \mathbf{U}_{t} = [\mathbf{u}_{0,t}\ \mathbf{u}_{1,t}\hdots \mathbf{u}_{K,t}],
    \end{align}
     and can be obtained using standard I-CSI-based precoders as illustrated in Section \ref{section_3}-B. Meanwhile, the power for each user is computed after randomly assigning a target SINR to the URLLC user as described next. In this specific case, the URLLC target SINR is drawn from a uniform distribution, \textit{i.e.},
\begin{align}\label{rand_power}
    \gamma_{0,t}^{tar} \sim \mathcal{U}\big(\gamma_{0,min}, \gamma_{0,max}\big),
    % p_{0,t} \sim \mathcal{U}\big(0, p_{max}\big],
\end{align}
where $\gamma_{0,min} = 2^{r_0}-1$ with $r_0$ as the spectral efficiency in bps/Hz, and $\gamma_{0,max}$ depicts the maximum theoretically achievable SINR, which corresponds to that obtained in an interference-free setup where all the power is allocated to the URLLC link and the BS uses MRT precoding.  Then, the system of equations
\begin{gather}\label{syst_eq}
 \begin{bmatrix}
 p_{0,t}|\mathbf{h}_{0,t}^H\mathbf{u}_{0,t}|^2 \!-\! \gamma_{0,t}^{tar}\sum\limits_{i\ne 0} p_{i,t}|\mathbf{h}_{0,t}^H\mathbf{u}_{i,t}|^2\\
 p_{1,t}|\mathbf{h}_{1}^H\mathbf{u}_{1,t}|^2 \!-\! \gamma_{1}^{tar}\sum\limits_{i\ne 1} p_{i,t}|\mathbf{h}_{1}^H\mathbf{u}_{i,t}|^2\\ 
 \vdots\\
 p_{K,t}|\mathbf{h}_{\scaleto{K}{4pt}}^H\mathbf{u}_{\scaleto{K}{4pt},t}|^2\!-\!\gamma_{\scaleto{K}{4pt}}^{tar}\sum\limits_{i\ne \scaleto{K}{4pt}} p_{i,t}|\mathbf{h}_{\scaleto{K}{4pt}}^H\mathbf{u}_{i,t}|^2
 \end{bmatrix}
 \!\!=\!\! 
  \begin{bmatrix}
  \gamma_{0,t}^{tar} \sigma^2\\
 \gamma_{1}^{tar} \sigma^2\\
 \vdots\\
 \gamma_{\scaleto{K}{4pt}}^{tar} \sigma^2
   \end{bmatrix}
\end{gather}
must hold in order to satisfy the SINR requirements of all the users. Hence, the power of each user can be obtained from solving \eqref{syst_eq}. After this step, we must check that the power allocation does not exceed the maximum available power $p_{max}$. If the power allocation is infeasible, \textit{i.e.,}  $\sum_{k = 0}^K p_{k,t}> p_{max}$, we draw a new URLLC SINR target sample according to \eqref{rand_power}, compute the users' power allocation that satisfies \eqref{syst_eq}, and repeat this process until the power constraint is fulfilled

After the power allocation, the $t-$th precoding matrix is formed as  
\begin{align}\label{W_structure}
     \mathbf{W}_{t} = [\sqrt{p_{0,t}}\mathbf{u}_{0,t}\ \sqrt{p_{1,t}}\mathbf{u}_{1,t}\hdots \sqrt{p_{K,t}}\mathbf{u}_{K,t}].
    \end{align}
    Then, the column vectors of matrix $\mathbf{W}_t$ are substituted into \eqref{eq2} and the SINR is computed for each of the $L$ past channel measurements of the URLLC user. Next, we compute $\mu_{UB}$ according to \eqref{eq6}. Finally, the precoding matrix that minimizes the transmit power while satisfying constraint \eqref{P2:b} constitutes the solution. Mathematically, the optimum index $t$ is obtained by solving
\begin{subequations}\label{P3}
	\begin{alignat}{2}
	\mathbf{P3:}\ \ t^{opt} = & \argminA_{t\in[1,\zeta]}  ||\mathbf{W}_{t}||_F^2,\label{P3:a}\\
	& \text{subject to} \ \ \ \mu_{UB}(\mathbf{W}_t) \le \xi. \label{P3:b}
	\end{alignat}
\end{subequations}
Observe that $\mathbf{P3}$ is always feasible given sufficiently large values of $\zeta$ and $r$. To prove it, we must consider the limit case when $r \to \infty$ and $\zeta \to \infty$. Under these conditions, there will always be at least a vector $\mathbf{h}_{0,t}$ from the set $\zeta$ whose associated precoders $\mathbf{w}_k$ ensure that $\gamma_{0,j}(\mathbf{w}_k)\ge \gamma_{0}^{tar}, \ \forall j$ since the image of $\gamma_{0,j}$ is the entire non-negative real domain. Now, from \eqref{eq6} and \eqref{P3:b}, we have 
\begin{align}\label{proof}
     \mu_{UB}=\hat{\mu}-\frac{\hat{s}}{\sqrt{L}}F_{\varphi}^{-1}(1-\alpha)&\leq \xi,
\end{align}
with $0\le\xi\le 1$. Now, computing the limits
\begin{align}
&\lim_{r \to \infty} \hat{\mu} = 0, \ \ \
\lim_{r \to \infty} \hat{s} = 0,\ \ \
\lim_{r \to \infty} \varphi =  -\mu.
\end{align}
Therefore, $F_{\varphi}^{-1}(1-\alpha)$ converges to $1-F_{\mu}^{-1}(1-\alpha)$, and $\mu_{UB}$ to zero. Then, \eqref{proof} becomes true independently of the reliability requirement $\xi$. This implies that for large values of $\zeta$ and $r$, there will be at least one solution $\mathbf{W}_t$ for $\mathbf{P3}$. It is worth noting that there is a trade-off on the selection of $\zeta$. On the one hand, very large values of $\zeta$ imply large processing times, which is not suitable for practical systems. On the other hand, very small values of $\zeta$ may not guarantee to find a feasible solution for the problem.

\begin{algorithm}[t!]\label{Algorithm}
\caption{Multi-antenna precoding for URLLC and eMBB coexistence}\label{algorithm}
\hspace*{\algorithmicindent} \textbf{Inputs:} $r,\ \{\tilde{\textbf{h}}_{0,j}\},\ \{\textbf{h}_k\}$  \\
\hspace*{\algorithmicindent}
\textbf{Outputs:} $\ \{\textbf{w}_k\}$
\begin{algorithmic}[1]
\State $p_T$ $\gets p_{max}$
\State Compute $\bar{\textbf{m}}$ and $\textbf{C}$ according to (10), (11)
\For{$t  =  1$ to $\zeta$}
\State Generate $\mathbf{h}_{0,t}$ with (12) 
\State Compute $\textbf{H}_t$ and $\textbf{U}_t$ according to (13), (14)
\State Draw $\gamma_{0,t}^{tar}$  according to \eqref{rand_power} 
\State Compute $p_{k,t}$ according to \eqref{syst_eq}
\If{$\sum_{k = 0}^K p_{k,t}\le p_{max}$}
\State compute $\mathbf{W}_t$ according to \eqref{W_structure}
\State compute $\mu_{UB}$ according to \eqref{eq6}
\If{$\mu_{UB}\le\xi \ \textbf{and}$ \ $||\mathbf{W}_t||_F^2$ $<$ $p_T$} 
    \State 
        $\mathbf{W}^{opt} \gets \mathbf{W}_t$
     \State $p_T$ $\gets ||\mathbf{W}_t||_F^2$
\EndIf
\Else
\State go to step 6
\EndIf
\EndFor
\end{algorithmic}
\end{algorithm}

Notice that the variable $r$ controls the tightness of the Chernoff bound. As previously stated, $\gamma_{0,j}(\{\mathbf{w}_k\})>\gamma_{0}^{tar}$ must be usually satisfied, thus, if very large values of $r$ are used, the empirical distribution of $e^{r(\gamma_{0}^{tar}-\gamma_{0,j}(\{\mathbf{w}_{k}\}))}$ gets farther from a Gaussian distribution, therefore, slowing the convergence of $\hat{\mu}$ to $\mu$. Consequently, this may also lead to the selection of precoders that do not ensure the reliability target in practice. Therefore, the  value of $r$ must be controlled to guarantee the expected results. Relatively small values, \textit{e.g.}, $r<10$, are suitable as discussed in Section \ref{section_4}.

Noteworthy, the precoding solutions obtained with our proposal ensure the reliability level as long as the number of channel measurements is large enough to compute the sample mean of the constraint in \eqref{P2:b} with confidence of $100\times \alpha\%$. 

\textbf{Algorithm 1} encloses the required steps for the solution of the initial problem given a certain $r$. The complexity of the algorithm is mainly dominated by the number of iterations and the solution of the system of equations (steps 3 and 7). The former increases the complexity by $\zeta$, while the latter by $(K+1)^3$, leading to a total complexity of $O(\zeta (K+1)^3)$.

\subsection{Precoding methods}
Herein, we consider two precoding schemes, ZF, and TPM\footnote{This precoder matches the structure of the optimal receive beamforming in the UL channel, \textit{i.e.,} MMSE, if we equate the parameters $\{\lambda_k\}$ to the UL transmit powers \cite{bjornson2014optimal}.}, and illustrate how to obtain the normalized precoders \eqref{U_structure} in such cases. Moreover, in case of ZF, we show how the procedure can be significantly simplified.
\subsubsection{ZF}
Under ZF, the links become noise-limited since the interference term is removed. The normalized ZF precoding vector is given by $\mathbf{u}_k= \mathbf{z}_k/||\mathbf{z}_k||$ with
\begin{equation}\label{zf_matrix}
    [\mathbf{z}_1\hdots\mathbf{z}_{K}] = \mathbf{H}(\mathbf{H}^H\mathbf{H})^{-1},
\end{equation}
where $\mathbf{H} = [\mathbf{h}_1 \hdots \mathbf{h}_K]$ depicts a matrix containing all instantaneous channel column vectors from the BS to all users. Notice that ZF requires the computation of the pseudo-inverse of a $K\times K$ matrix which might be computationally costly. Fortunately, the computation complexity decreases as $M$ grows as in massive MIMO systems since the term $(\mathbf{H}^H\mathbf{H})^{-1}/M$ converges to the identity matrix. In such asymptotic regime, the expression in \eqref{zf_matrix} becomes $[\mathbf{z}_1\hdots\mathbf{z}_{K}] = \mathbf{H}$, which matches the MRT precoding \cite{khandaker2018signal}.

Back to the proposed algorithm, for each $\mathbf{H}_t$ \eqref{h_matrix}, one evaluates expression \eqref{U_structure} as
\begin{align}
\mathbf{U}_{t} = \bigg[\frac{\mathbf{z}_{0,t}}{||\mathbf{z}_{0,t}||}\ \frac{\mathbf{z}_{1,t}}{||\mathbf{z}_{1,t}||}\hdots  \frac{\mathbf{z}_{K,t}}{||\mathbf{z}_{K,t}||}\bigg],
\end{align}
where
\begin{align}
    [\mathbf{z}_{0,t} \ \mathbf{z}_{1,t} \hdots \mathbf{z}_{K,t}] = \mathbf{H}_t(\mathbf{H}_t^H\mathbf{H}_t)^{-1}.
\end{align}
With ZF, the SINR expression in \eqref{eq2} reduces to  
\begin{align}\label{sinr_embb}
    \gamma_{k}(\{\mathbf{w}_{k,t}\}) \!&=\! \frac{|\mathbf{h}_{k}^H\mathbf{w}_{k,t}|^2}{\sigma^2} \!=\! \frac{p_{k,t}}{||\mathbf{z}_{k,t}||^2\sigma^2}.
\end{align}
Therefore, to determine the power allocation required to find \eqref{W_structure}, we perform \eqref{rand_power} and isolate the powers $p_{k,t}$ from \eqref{sinr_embb} after accordingly replacing    $\gamma_{k}(\{\mathbf{w}_{k,t}\})$ by $\gamma_{0,t}^{tar}$ and $\gamma_{k}^{tar}$. Notice that the above SINR expression only contains one variable $p_{k,t}$, thus, solving the system of equations in \eqref{syst_eq} can be avoided. Finally, one must proceed to determine $\mathbf{W}_t$ in \eqref{W_structure} and $\mu_{UB}$ according to  \eqref{eq6} to then choose the precoder with the minimum allocated power, \textit{i.e.,} the solution of  $\mathbf{P3}$.
\subsubsection{TPM}
 The normalized TPM precoding vector is given by \cite{bjornson2014optimal}
 \begin{align}\label{u_TPM}
     \mathbf{u}_k = \cfrac{\bigg(\mathbf{I}_M + \sum_{i = 0}^K\cfrac{\lambda_i}{\sigma^2}\mathbf{h}_i\mathbf{h}_i^H\bigg)^{-1}\mathbf{h}_k}{\bigg\lVert\bigg(\mathbf{I}_M + \sum_{i = 0}^K\cfrac{\lambda_i}{\sigma^2}\mathbf{h}_i\mathbf{h}_i^H\bigg)^{-1}\mathbf{h}_k\bigg\rVert},
 \end{align}
 where $\mathbf{I}_M$ depicts the identity matrix and $\{\lambda_i\}$ represent the Lagrange multipliers used to solve the original problem in $\mathbf{P1}$ without the URLLC constraint. The latter can be computed from fixed-point equations as
 \begin{align}\label{lambdas}
     \lambda_k = \cfrac{\sigma^2}{\Big(1+\cfrac{1}{\gamma_k}\Big)\mathbf{h}_k^H\Big(\mathbf{I}_M+\sum_{i = 0}^K\cfrac{\lambda_i}{\sigma^2}\mathbf{h}_i\mathbf{h}_i^H\Big)^{-1}\mathbf{h}_k}.
 \end{align}
Given the above precoding structure, we can perform \eqref{sample_mean}-\eqref{h_matrix}, and then use \eqref{lambdas}-\eqref{u_TPM} to compute $\mathbf{U}_t$. The next step is the random SINR assignment \eqref{rand_power}.  Notice that with TPM, the interference term is not perfectly removed as in ZF. Therefore, to compute the power allocation $p_{k,t}$, one must unavoidably find the solution of the system of equations in \eqref{syst_eq}. Again, to conclude, one must compute the precoding matrix $\mathbf{W}_t$ given that $\mathbf{w}_t = \sqrt{p_{k,t}}\mathbf{u}_k$, determine  $\mu_{UB}$ according to  \eqref{eq6}, and  find the solution to $\mathbf{P3}$.
 
 \subsection{Practicalities}

 To reduce the processing delay, the computations related to the proposed algorithm may be executed in instances where only the eMBB users are receiving data. Notice that during this time, the BS will not use the precoders obtained from the proposed algorithm for current DL transmissions, but the ones computed in parallel using predefined precoding schemes. In general, the algorithm should run every time the I-CSI of eMBB users is updated. Meanwhile, to reduce the hardware resource utilization, it might not be necessary to run the algorithm every time a new URLLC channel measurement is obtained, since the statistics will not be considerably modified.

The proposed algorithm also considers a single URLLC transmission per resource block. This is because concurrent URLLC transmissions would cause a strong mutual interference due to the use of imperfect statistics on the precoder design. Notice that we refer here to URLLC users with very tight latency requirements that do not allow I-CSI acquisition. Other URLLC services with not that stringent latency demands might be treated in the same way as eMBB users.

Finally, the  storage space required to save the channel measurement will be upper bounded by $\sum_{\forall u}10/\xi_u\times B \times M$ with $\xi_u$ and $B$ as the user-specific outage target and the number of bits required for quantization, respectively. For instance, for 10 URLLC users, $\xi_u = 10^{-3}$, $B=8$, and  $M = 8$, the maximum storage space that would be required is 6400000 bits (0.8 MB). Notice that a finite quantization level may affect the performance in practice, which could be considered in future works.

\begin{table}[t!]
    \centering
    \caption{Simulation parameters}
    \label{table_2}
    \begin{tabular}{c c c c}
        \hline
        \textbf{Parameter} & \textbf{Value}& \textbf{Parameter} & \textbf{Value} \\
            \hline           
            $M$& \{8, 16\} & $K$ & 4 \\
            $\delta$& 3.5 & $p_{max}$ & 47 dBm \\
             $d_r$&  500 m & $\gamma_{0}^{tar}$ & -11.44 dB \\
            $\kappa_0$ & \{0, 2, 5, 10\} & $\gamma_{k}^{tar}(k\ne 0)$ & \{0, 10\} dB  \\
            $r$& 10 & $\xi$ & $\{10^{-3}, 10^{-4}\}$ \\
             $L$& 250, 500, 3500 & $\alpha$ & 0.99 \\
             $\kappa_k  (k\ne 0)$  &  $\{0, 2\}$ &  $\zeta$ & 3000\\
            \hline
    \end{tabular}
\end{table}
\section{Numerical results}\label{section_4}
We consider that the BS serves one URLLC and four eMBB users within a resource block. 
The users are uniformly deployed in an area of radius $d_r$ around the BS. The distance between the BS and user $k$ is denoted as $d_k$, while the path gain experienced by the latter is given by $\psi_k = d_k^{-\delta}$, where  $\delta$ depicts the path loss exponent. Moreover, we use the Rician fading model due to its potential to cover different scenarios by properly tuning the parameter $\kappa_k$, from non LOS (NLOS) setups as in Rayleigh fading ($\kappa_k = 0$) to fully deterministic LOS scenarios ($\kappa_k\xrightarrow[]{}\infty$). Specifically, the Rician channel model is given by \cite{hampton2013introduction}
\begin{align}
    \mathbf{h}_k = \sqrt{\psi_k}\bigg(\sqrt{\frac{\kappa_k}{\kappa_k+1}}\mathbf{h}_{k,LOS} + \sqrt{\frac{1}{\kappa_k+1}}\mathbf{h}_{k,NLOS}\bigg),
\end{align}
where $\psi_k\sqrt{\kappa_k/(\kappa_k+1)}\mathbf{h}_{k,LOS}$ represents the deterministic LOS propagation component, and $\psi_k\mathbf{h}_{k,NLOS}/\sqrt{\kappa_k+1}$ represents the scattering component with $\mathbf{h}_{k,NLOS}, \sim \mathcal{CN}(\mathbf{0},\mathbf{I})$. For simplicity, we set $\mathbf{h}_{k,LOS}$ as a vector of ones and assume the same $\gamma_{k}^{tar}$  and LOS factor $\kappa_k$ for all the eMMB users. The remaining simulation parameters are shown in Table~\ref{table_2}. Notice that the noise power corresponds to a bandwidth of 10 MHz, while $\gamma_{0}^{tar}$ comes from assuming a packet of 32 bytes transmitted over 0.256 ms and 10 MHz, i.e., $\gamma_{0}^{tar}=2^{32\times 8 \text{bits}/(0.256\times10^{-3} \text{s}\times 10^7 \text{Hz})}-1= 0.0718 = -11.44$ dB.

In Sections \ref{section_4}-A, \ref{section_4}-B and \ref{section_4}-C, we evaluate the performance of the proposed algorithm for an instantaneous network realization, including a given URLLC channel history, network deployment, and I-CSI of the eMBB users.\footnote{Given a certain seed for the generation of random numbers, we obtained a single network realization. We repeatedly tested many seeds and verified that the performance trends remain similar.} Meanwhile, we present statistics obtained over $5\times10^3$ network realizations in Section~\ref{section_4}-D.
\begin{figure}[t!]
    \centering
    \includegraphics[width = 0.9\columnwidth]{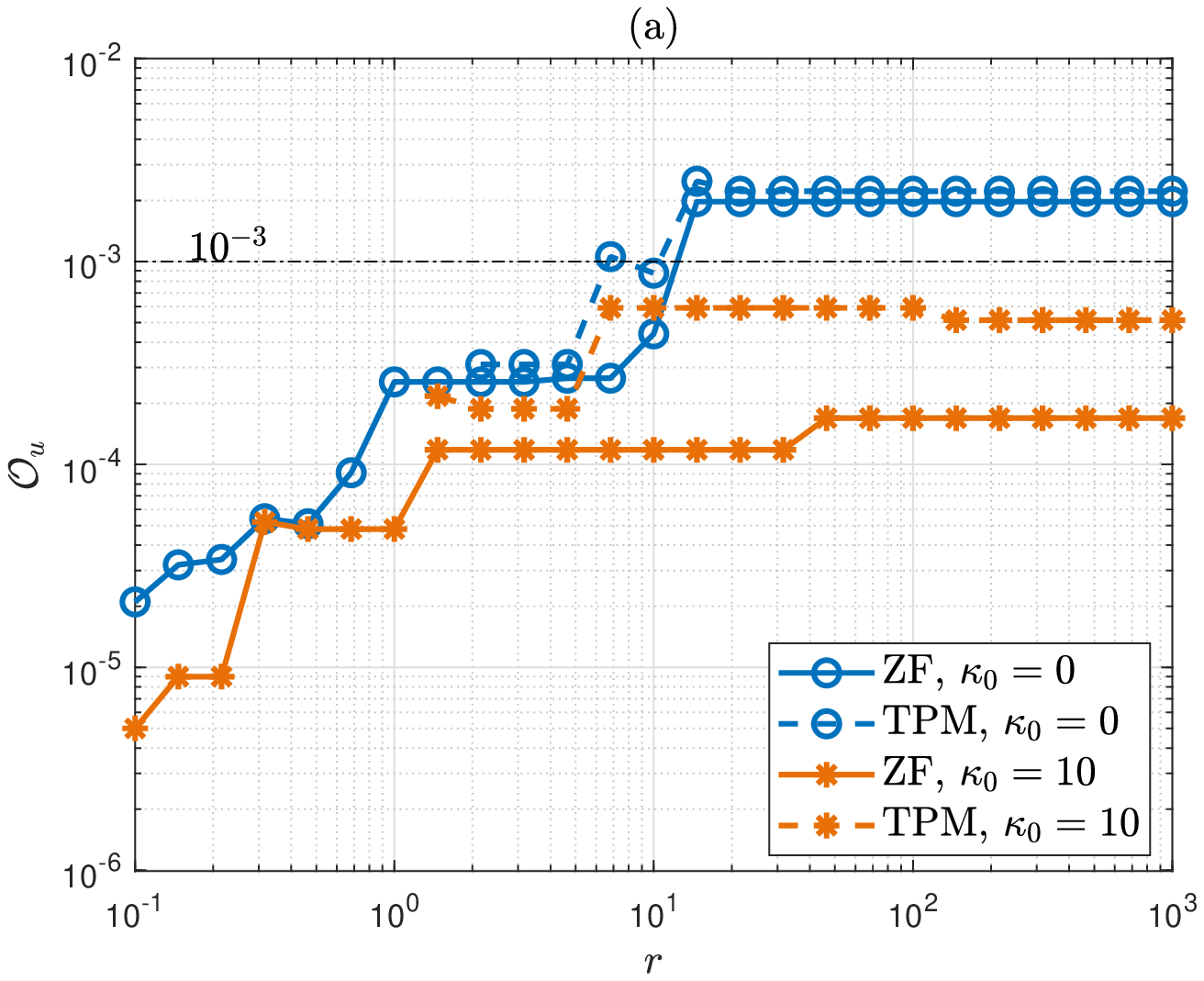}
    \includegraphics[width = 0.9\columnwidth]{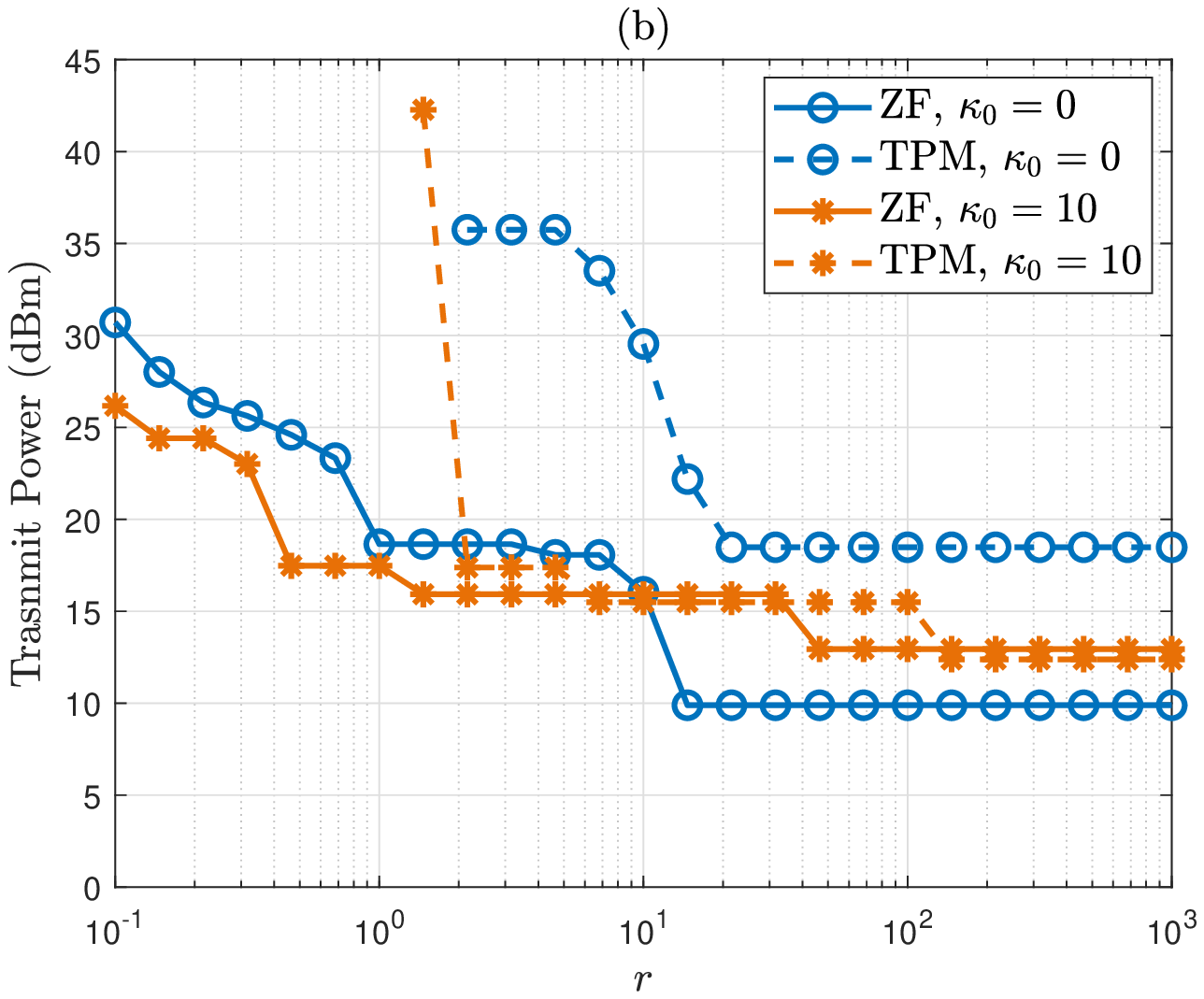}
    \caption{(a) Outage probability, and (b) total transmit, power as a function of $r$. The URLLC channel is subject to Rayleigh fading ($\kappa_0 = 0$) and Rician fading ($\kappa_0 = 10$), while $\xi = 10^{-3}$, $\gamma_k^{tar}= 10$ dB $\forall k \ne 0$, and $L=500$ channel measurements.}
    \label{fig:2}
\end{figure}
\subsection{
On the configuration of $r$ and $\zeta$}
 Fig. \ref{fig:2} shows the impact of  the configuration of the parameter $r$ on the attainable outage probability (Fig. \ref{fig:2} (a)) and transmit power (Fig. \ref{fig:2}(b)) for the proposed algorithm with both ZF- and TPM-based precoding. Herein, we set the outage target to $10^{-3}$, and assume $L = 500$ channel measurements. Notice that the outage probability decreases as $r$ decreases, but at the cost of attaining higher-power precoders. Meanwhile, adopting relatively large values of $r$ leads to reduced transmit power precoders that may not satisfy the outage constraint. Interestingly,  ZF outperforms TPM in terms of transmit power regardless of the value of $r$. Noteworthy, the TPM precoding is the optimal for transmit power minimization only with I-CSI availability. Nevertheless, the increment of $\kappa_0$ brings a reduction in the performance gap between both precoding methods since the channel becomes more deterministic and the randomly generated coefficients are closer to the actual channel realizations. Notice that the increment of $\kappa_0$  also eases the selection of $r$. In the following,  we set $r = 10$, which allows reliably meeting the outage constraint for both ZF and TPM-based precoding mechanisms as illustrated in Fig. \ref{fig:2}.
 
\begin{figure}[t!]
    \centering
    \includegraphics[width = 0.9\columnwidth]{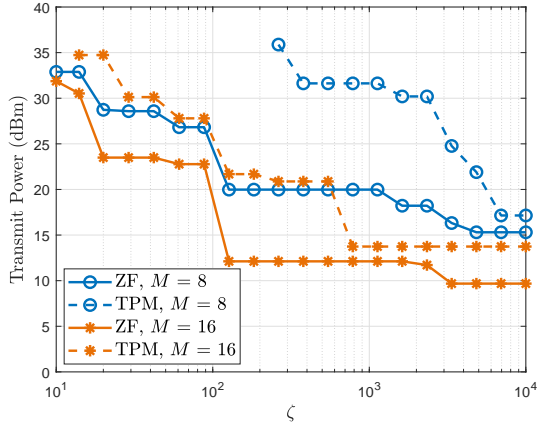}
    \caption{Total transmit power as a function of $\zeta$. The URLLC channel is subject to Rayleigh fading ($\kappa_0 = 0$), while $\xi= 10^{-3}$, $\gamma_k^{tar}= 10$ dB $\forall k \ne 0$, and $L=500$ channel measurements.}
    \label{fig:6}
\end{figure}
\begin{figure}[t!]
    \centering
    \includegraphics[width = 0.9\columnwidth]{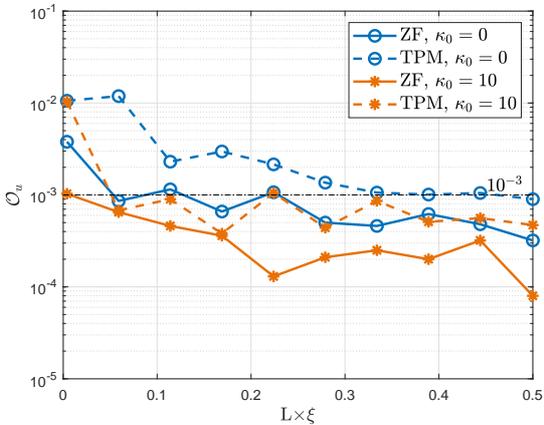}
    \caption{Outage probability vs L$\times\xi$ for Rayleigh ($\kappa_0 = 0$) and Rician ($\kappa_0 = 10$) fading scenarios for ZF and TPM with $\gamma_{k}^{tar} = 10$~dB $\forall k\neq 0$.}
    \label{fig:3}
\end{figure}

Fig. \ref{fig:6} shows the impact of the number of random generated vectors on the total transmit power for both precoding schemes. Observe that as $\zeta$ increases, the obtained precoder gets probabilistically closer to the optimum one. Interestingly, the transmit power reduction is not significant for $\zeta>3000$ in most configurations. Therefore, adopting higher values is not convenient as it increases the processing times. Observe for a relatively small $\zeta$, the problem may not be feasible, as it occurs for $\zeta<280$ for TPM and $\zeta<12$ for ZF, with $M=8$. However, the problem becomes feasible as $\zeta$ gets relatively larger, which confirms the statement in Section~\ref{section_3}-A about the feasibility of the problem as $\zeta\to\infty$.
\subsection{On the performance impact of the number of measurements and URLLC LOS channel factor}
\begin{figure}[t!]
    \centering
    \includegraphics[width = 0.9\columnwidth]{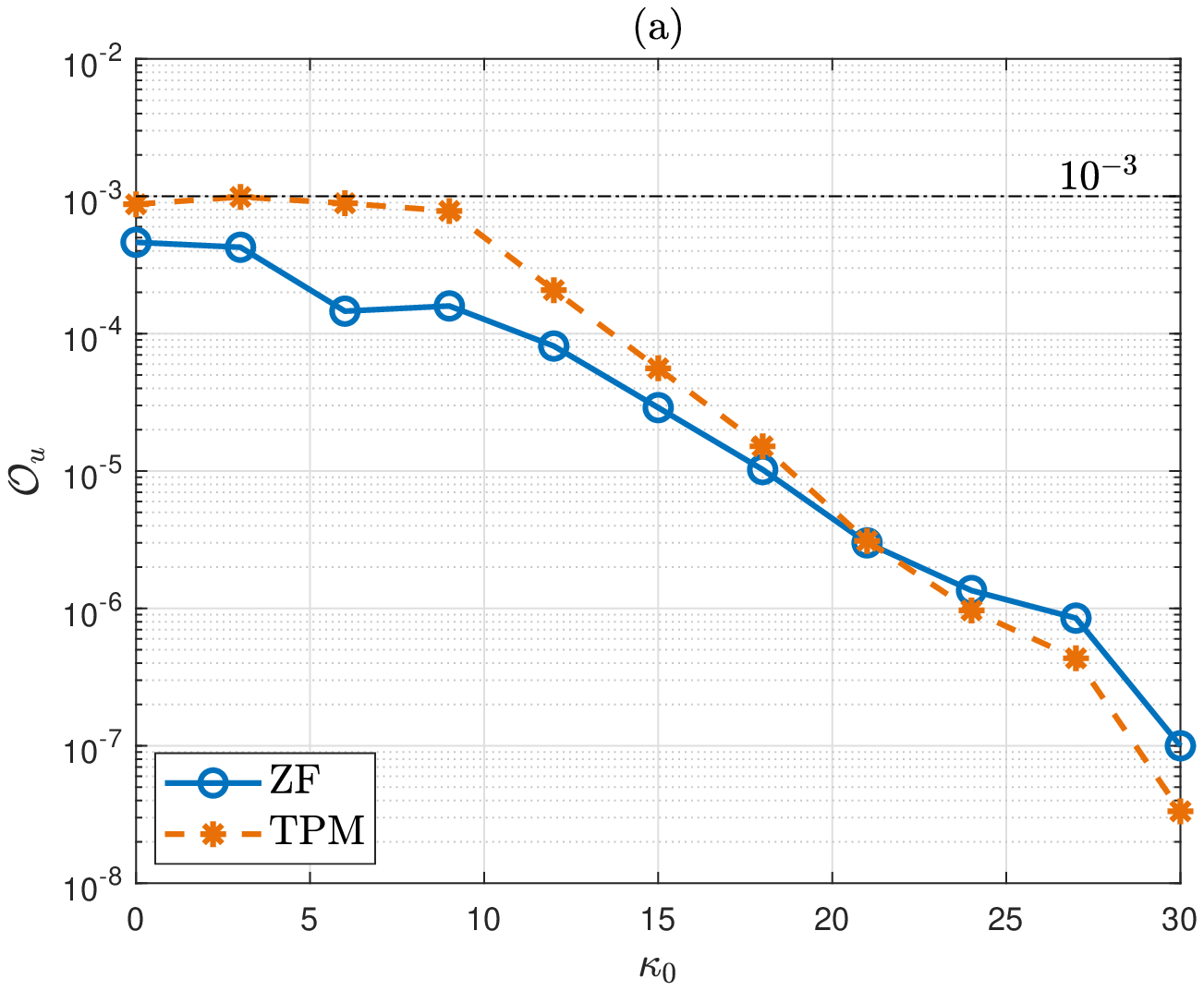}
    \includegraphics[width = 0.9\columnwidth]{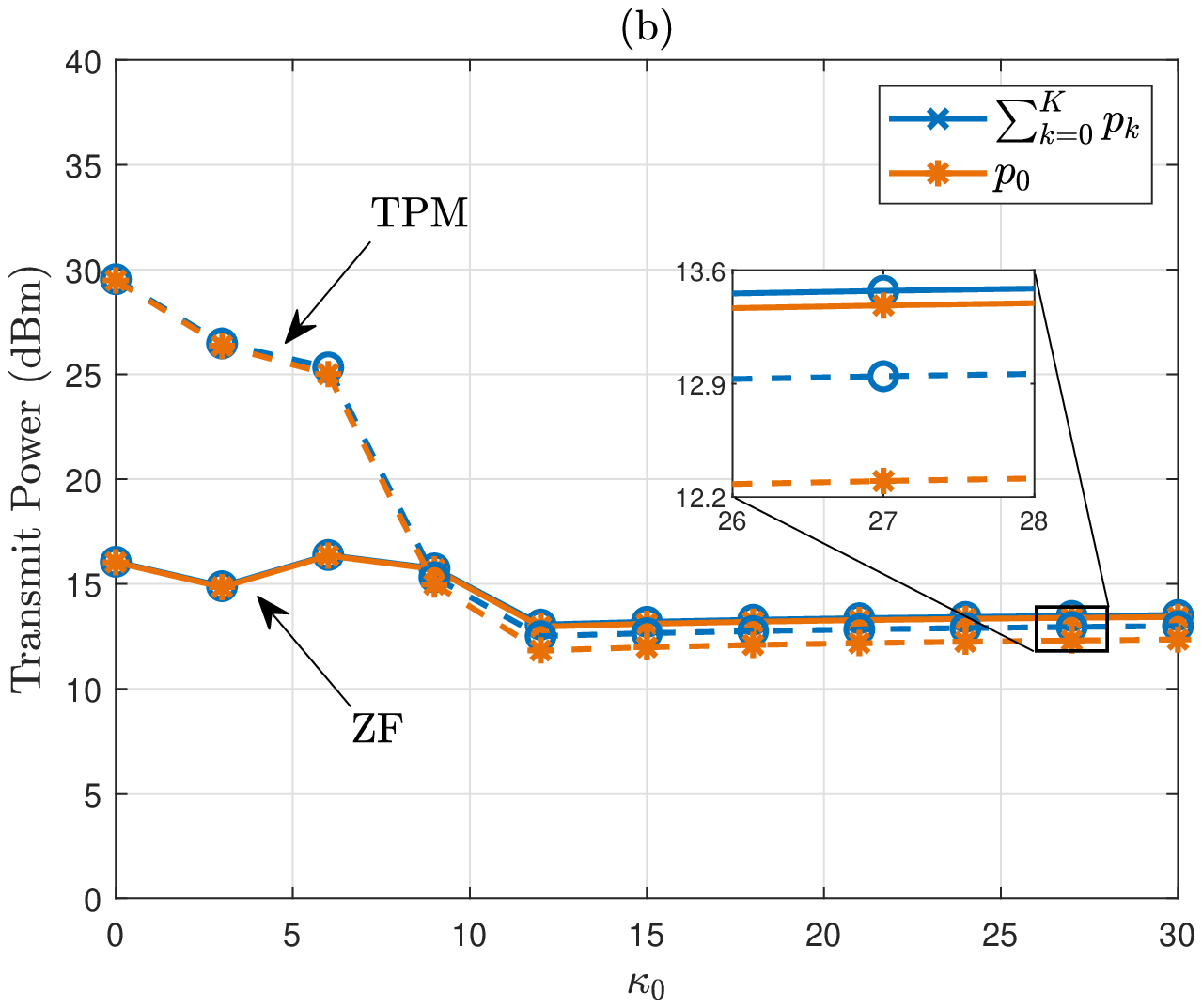}
    \caption{(a) Outage probability, and (b) total transmit power and transmit power for URLLC as a function of $\kappa_0$. We set $\gamma_{k}^{tar} = 10 $~dB $\forall k\ne 0$, $\xi = 10^{-3}$, and consider $L=500$ channel measurements.}
    \label{fig:4}
\end{figure}

Fig. \ref{fig:3} shows the minimum number of channel measurements required to keep the outage target below the threshold $\xi$. For ZF, around 250 past channel measurements are enough to achieve the target $\xi=10^{-3}$ when $\kappa_0 = 0$, \textit{i.e.,} Rayleigh fading, while the number drops below 50 when $\kappa_0 = 10$. However, these figures considerably increase for TPM, being around 470 and 120, respectively. The reduction of the required number of samples as $\kappa_0$ increases is because the generated channel vectors get more concentrated around the actual (more deterministic) channel realizations.

 Fig. \ref{fig:4} (a) displays the achievable outage probabilities for different values of $\kappa_0$. It is worth noting that the outage probabilities even go below $10^{-5}$ as $\kappa_0$ increases. This guarantees high reliability levels, even if the outage target is more stringent, in scenarios where the BS and users have a relatively strong LOS component. Meanwhile, Fig. \ref{fig:4} (b) shows the behavior of the transmit power for different values of the LOS component. The increment of $\kappa_0$ leads to lower transmit powers which converge due to the fixed number of random generations $\zeta$. The figure also depicts the fact that most of the power is required for the URRLC service, evincing the significant costs of achieving high reliability. Also notice that ZF outperforms TPM when $\kappa_0\le 8$, while TPM performs better as $\kappa_0$ increases since the generated coefficients are closer to the actual channel realizations of the URLLC user.

In the following, we focus only on the performance of ZF. This is for simplicity and given ZF outperforms TPM precoding in poor channel conditions. Moreover, we consider $L = 250$ channel measurements for which, given $r=10$, the URLLC constraint is already met as shown in Fig.~\ref{fig:3}.

\subsection{On the performance impact of the number of eMBB users and their SINR target}

\begin{figure}[t!]
    \centering
    \includegraphics[width = 0.9\columnwidth]{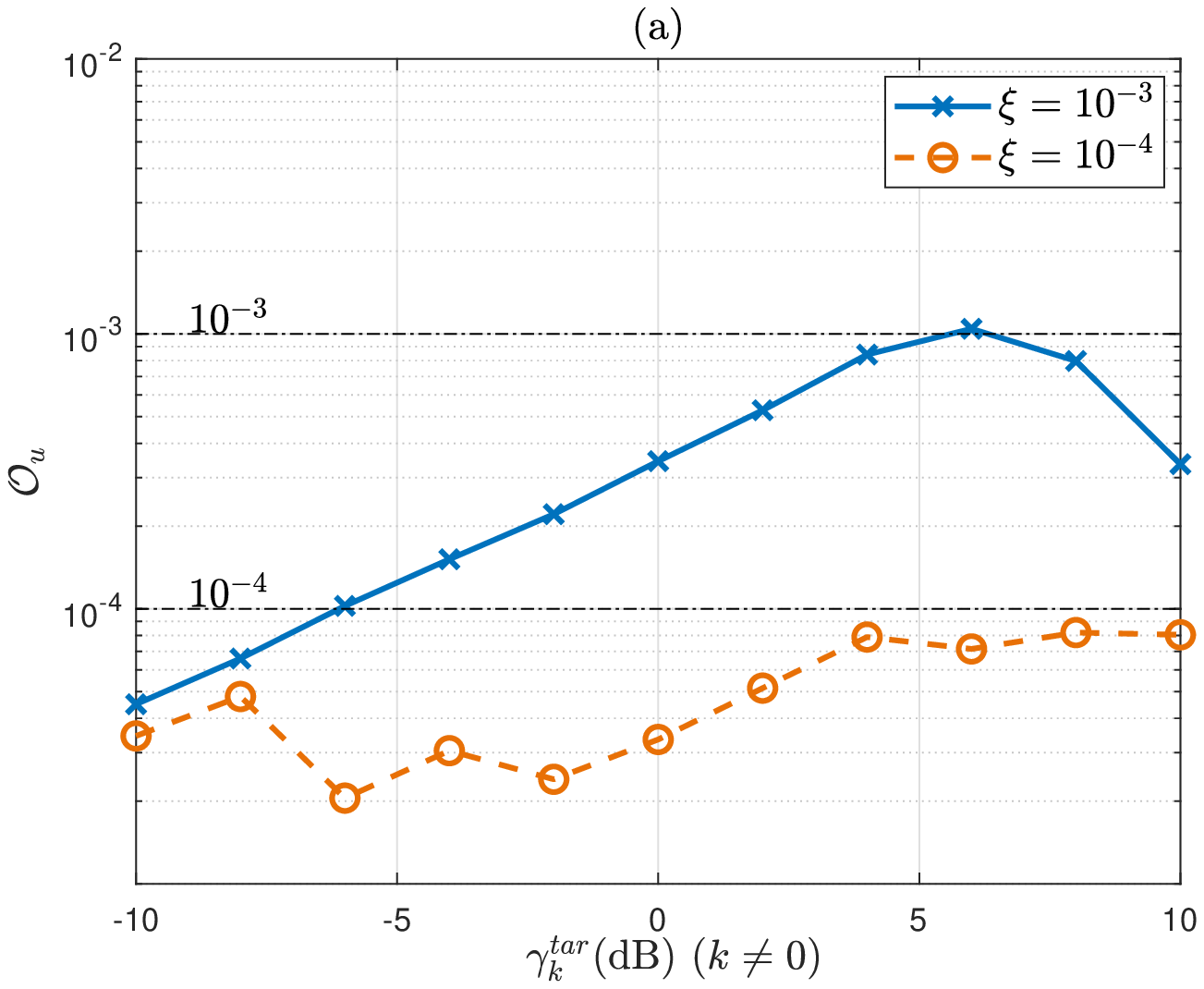}
    \includegraphics[width = 0.9\columnwidth]{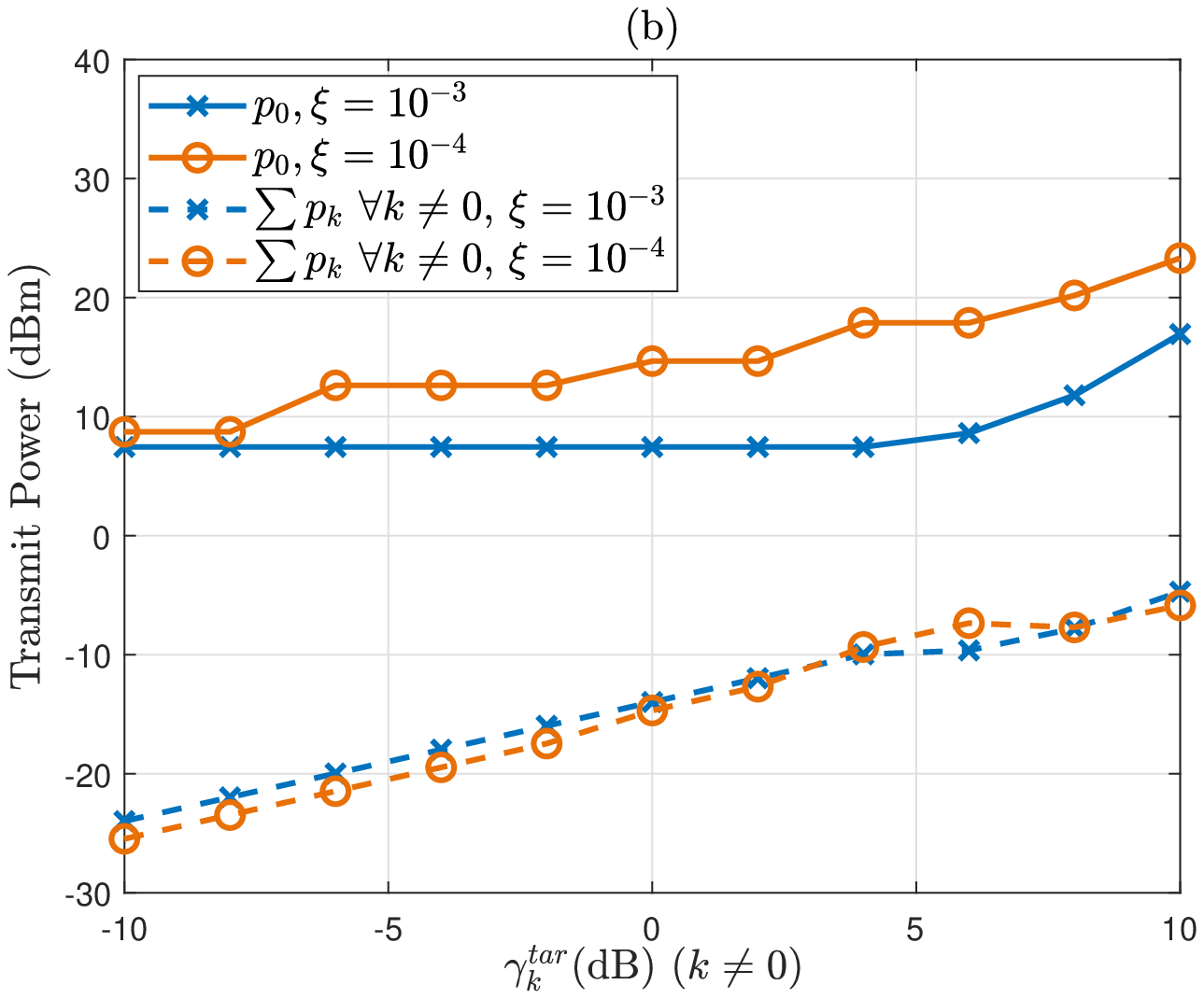}
    \caption{(a) Outage probability, and (b) transmit power of both, eMBB and URLLC, DL transmissions, as a function of $\gamma_{k}^{tar}$ ($k\ne 0$). URLLC channel is subject to Rayleigh fading ($\kappa_0 = 0$). We set $L = 250, 3500$ channel measurements for $\xi = 10^{-3}$, $10^{-4}$, respectively.}
    \label{fig:8}
\end{figure}
\begin{figure}[t!]
    \centering
    \includegraphics[width = 0.9\columnwidth]{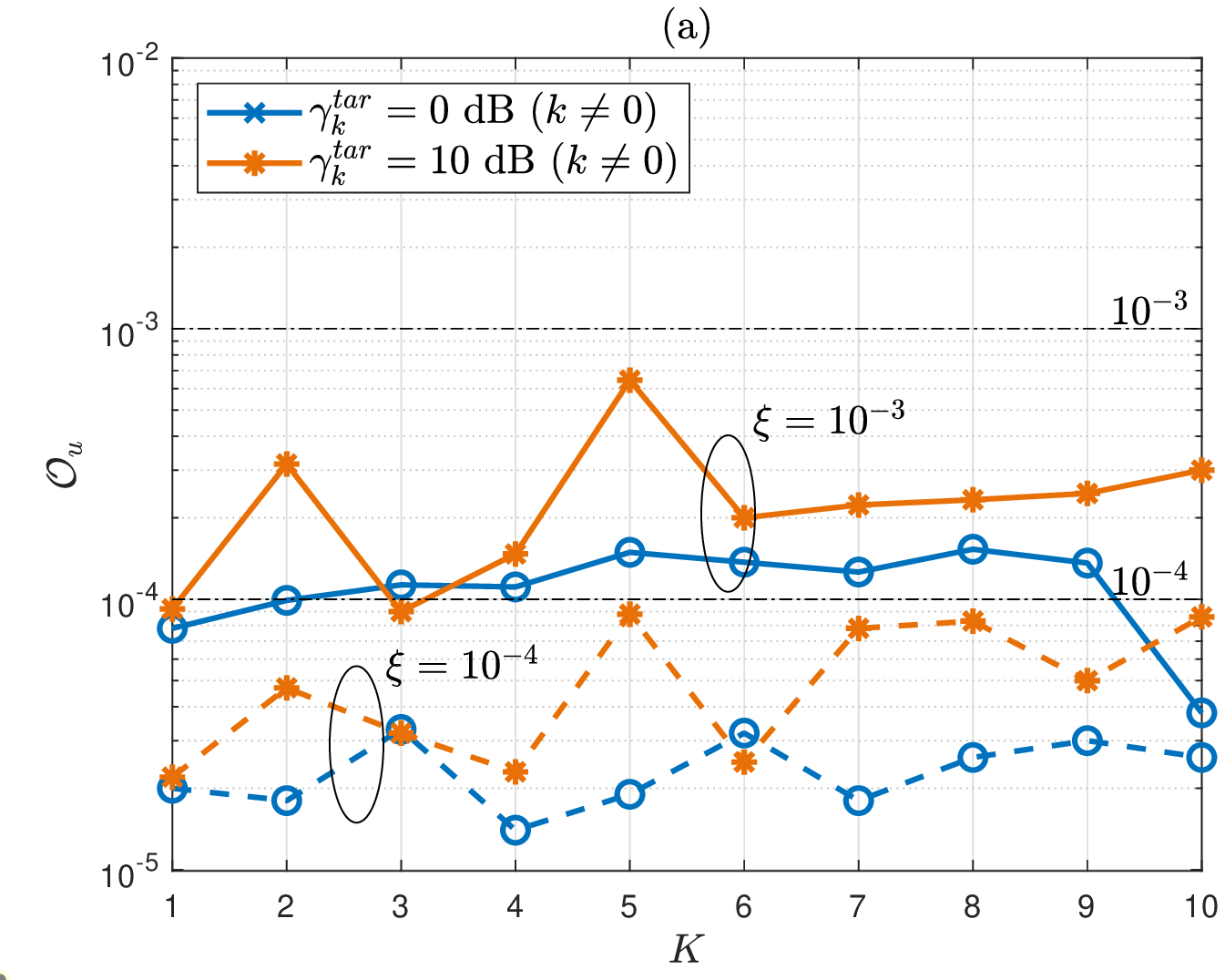}
    \includegraphics[width = 0.9\columnwidth]{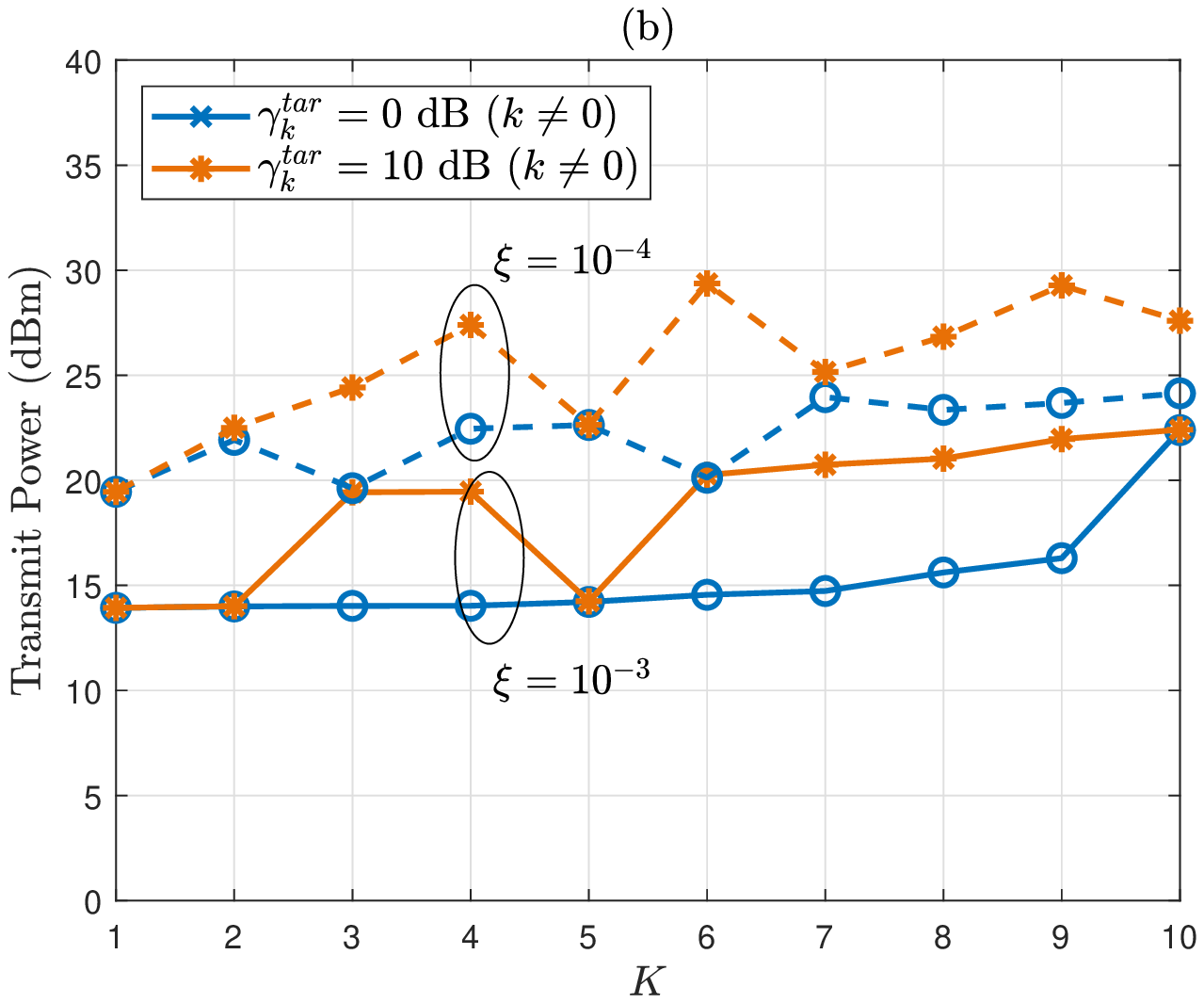}
    \caption{(a) Outage probability, and (b) total transmit power, as a function of the number of simultaneously served eMBB devices. The URLLC channel is subject to Rayleigh fading ($\kappa_0 = 0$). We set $M = 16$, and $L = 250,3500$ channel measurements for $\xi = 10^{-3}, 10^{-4}$, respectively.}
    \label{fig:nDevices}
\end{figure}
 
  Fig. \ref{fig:8} shows the behavior of the outage probability (Fig. \ref{fig:8} (a)) and transmit power (Fig. \ref{fig:8} (b)) for  different SINR targets $\gamma_k^{tar}$ $\forall k\ne 0$. Notice that the outage probabilities tend to increase with the SINR since larger eMBB transmission powers cause larger interference levels to the URRLC link. Meanwhile, higher SINR requirements of the eMMB users lead to more power allocated to them, therefore, increasing the total transmit power. Moreover, there is an increment on the transmit power intended to the URRLC user as tighter reliability targets are set. For instance, the required transmit power at $\gamma_{k}^{tar}$ = 10 dB $(k\ne 0)$ is approximately 17 dBm and 23 dBm, for $\xi = 10^{-3}$ and $\xi = 10^{-4}$, respectively. Notice that we are considering the worst  possible case (Rayleigh fading), where the  powers requirements are higher due to the lack of a LOS component. 

 Fig. \ref{fig:nDevices} (a) shows the achieved outage probability as a function of the number of eMBB devices that are simultaneously served within a resource block for $\xi = 10^{-3}$ and $\xi = 10^{-4}$. Note that the outage probability tends to increase with the number of devices, since the precoding needs to cope with larger interference levels. Therefore, higher transmit powers are required to achieve the targeted outage probabilities in the URLLC link, which is depicted in Fig. \ref{fig:nDevices} (b).

\subsection{On the statistics of the achievable outage probability and allocated transmit power}

\begin{figure}[t!]
    \centering
    \includegraphics[width = 0.9\columnwidth]{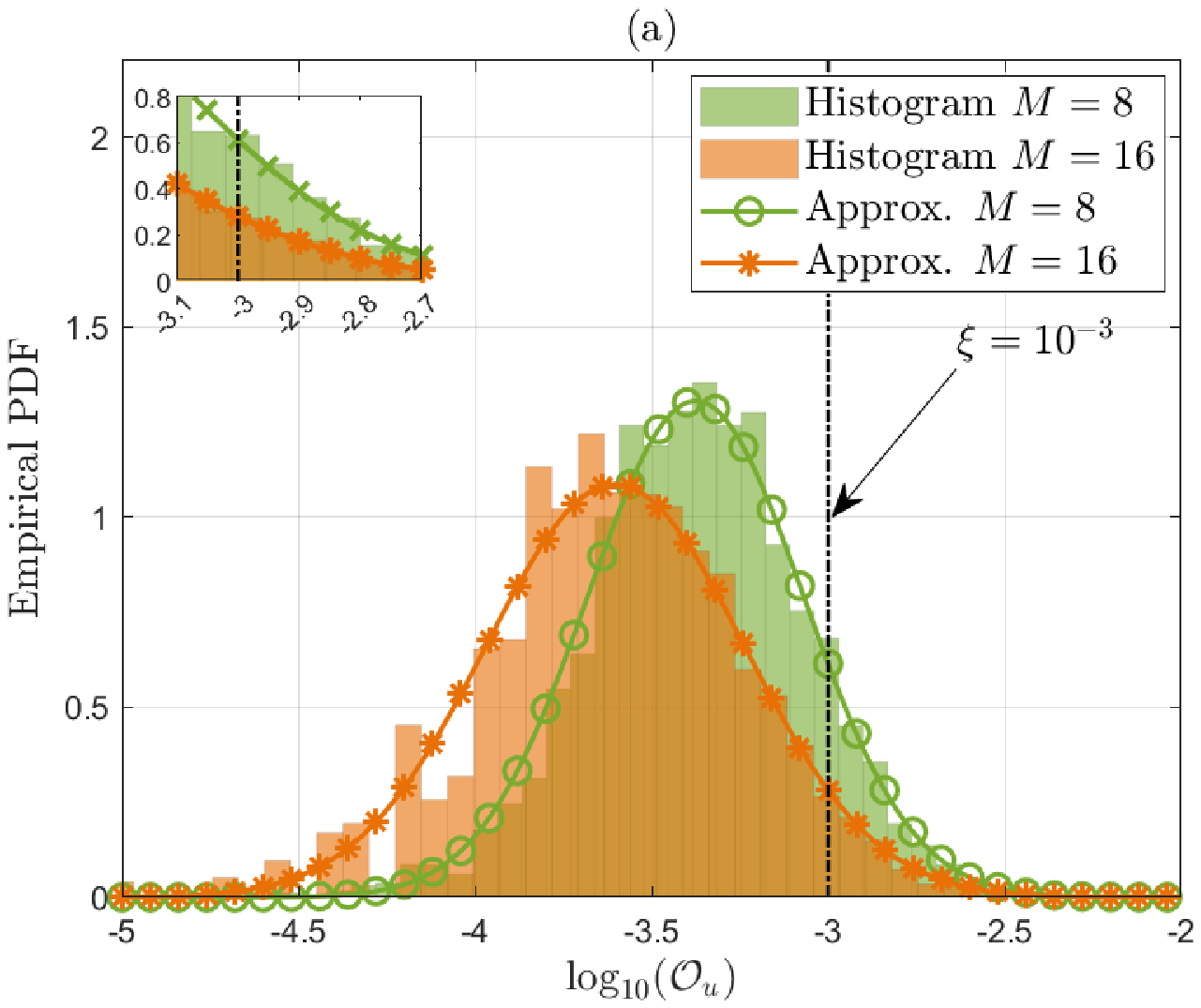}
    \includegraphics[width = 0.9\columnwidth]{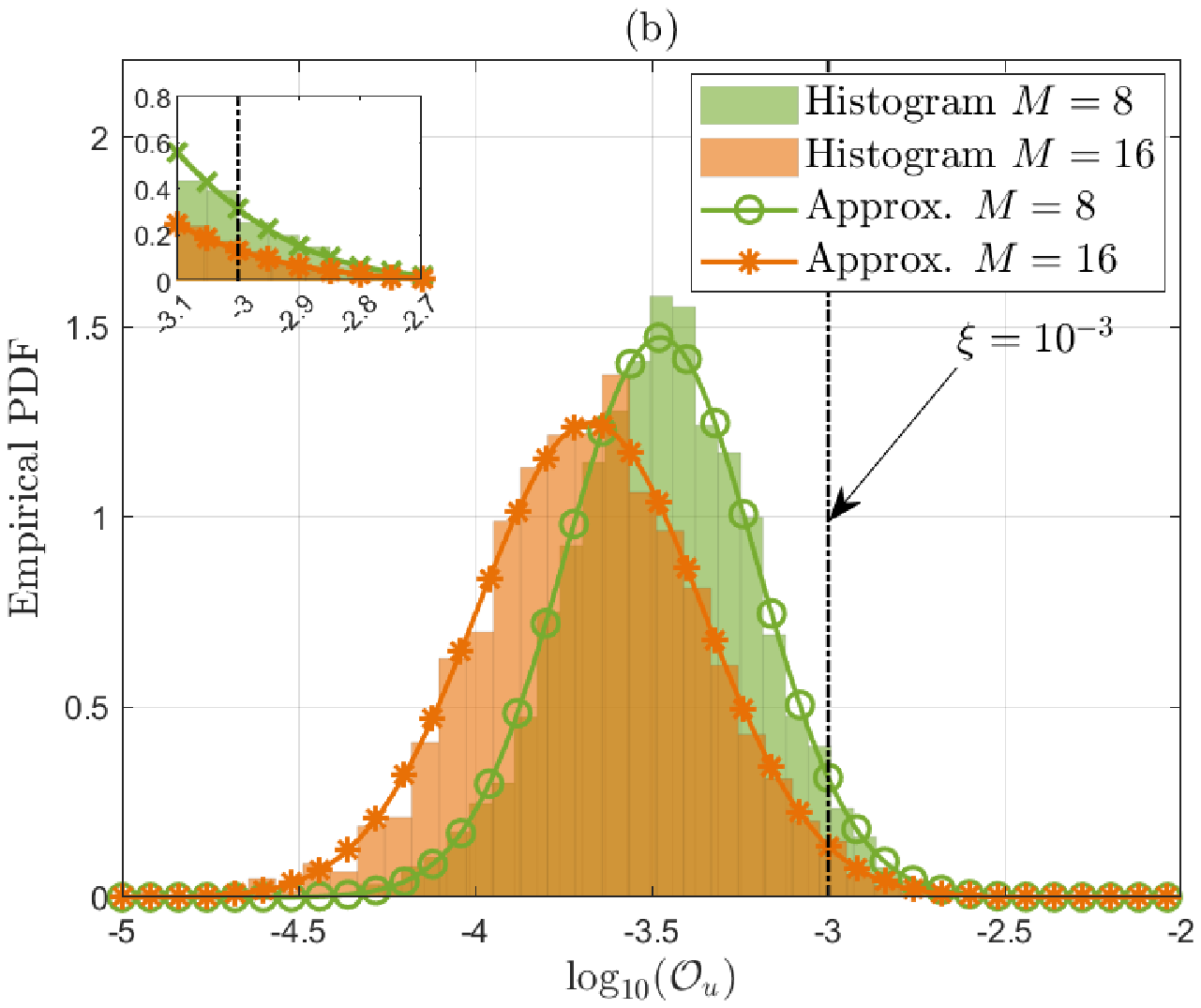}
    \caption{Empirical PDF of the outage probability and approximation to a Gaussian distribution for 5000 network realizations with (a) $L = 250$, and (b) $L = 500$ channel measurements. All users are subject to Rayleigh fading ($\kappa_k = 0 \ \forall k$). We set $\gamma_{k}^{tar} = 10$ dB $\forall k \ \ne 0$, and $\xi = 10^{-3}$}.
    \label{fig:10}
\end{figure}

 Different from the previous results, herein we obtain statistics for $5\times10^3$ randomly generated network realizations, \textit{i.e.,} different network deployments, channel history, and I-CSI of URLLC and eMBB users, respectively.

 Fig. \ref{fig:10} shows the empirical  probability density function (PDF) of the outage probability exponent, \textit{i.e.}, $\log_{10}\mathcal{O}_u$, for ZF with a target $\xi=10^{-3}$.
%with $M=8$ and $M=16$.
Notice that the histograms for  $L=250$ (Fig.~\ref{fig:10}~(a)) and $L=500$ (Fig.~\ref{fig:10}~(b)) approximately match a Gaussian PDF, whose parameters are obtained by standard curve fitting and are displayed in Table~\ref{table_3}. Here, the confidence value for $\xi \le 10^{-3}$ is obtained as
\begin{align}\label{confidence_eq}
    CV =\bigg(1-Q \bigg[\frac{\xi-MV}{SD}\bigg]\bigg)\times100,
\end{align}
where $MV$ depicts the estimated mean, and $SD$ the estimated standard deviation. In ZF precoding, we obtain $CV \approx 89.01\%$ and $CV \approx 94.96\%$ with $M = 8$ and $M = 16$, respectively, and exploiting $L=250$. Note that
the use of more antennas moves the mean of the distribution to the left due to the diversity gain. The improvement is, however, small because the algorithm reduces the transmit power while pushing $\mathcal{O}_u$ close to the target. It is worth highlighting that the  distributions can also move to the left in scenarios with larger $\kappa_0$ , and/or by exploiting more URLLC past channel measurements, e.g., $L=500$. Indeed, the chances of exceeding $\xi$ decrease considerably with the increment of $L$. Specifically, for $L=500$, the confidence levels increase up to $CV = 96.05\%$ and $CV = 98.25\%$ for $M=8$ and $M = 16$, respectively. Notice that the estimated values are close to the ones obtained with Monte Carlo (MC) simulations which means that a Gaussian distribution is a good approximation. The confidence levels can also be increased at the cost of incurring in higher transmit powers, \textit{i.e.,} smaller $\zeta$ or $r$. Meanwhile, TPM again exhibits a poor performance in Rayleigh fading, but it considerably improves as the value of $\kappa_0$ gets larger. The use of more more past channel measurements would strongly improve the $CV$ for this precoding method.  Also note that the confidence levels  for both precoding methods increase and even reach values above 99\% as the LOS components get stronger.

\begin{table}[t!]
    \centering
    \caption{Fitting parameters and confidence for $\log_{10}\mathcal{O}_u$}
    \label{table_3}
    \begin{tabular}{|c|c| c| c| c| c| c| c|}
        \hline
        Precod.&$\kappa_0$&$L$ &  $M$ & $MV$ & $SD$ & $CV(\%)$ & $MC(\%)$\\
            \hline           
     \parbox[t]{2mm}{\multirow{12}{*}{\rotatebox[origin=c]{90}{ZF}}}  &0 & 250 & 8 & -3.375 & 0.305 &
            89.01 & 89.50\\
             & 0& 250 & 16 & -3.603 & 0.367 & 94.96 & 95.46\\
             & 0& 500& 8 & -3.476 & 0.271 & 96.05 & 96.26\\
             & 0& 500 & 16 & -3.674 & 0.319 & 98.25 & 98.46 \\
              & 2& 250 & 8 & -3.929 & 0.604 &
            93.81 & 95.60\\
             & 2& 250 & 16 & -5.358 & 0.840 & 99.75 & $>$99.99\\
             & 2& 500& 8 & -4.729 & 0.815 & 98.30 & 99.40\\
             & 2& 500 & 16 & -6.643 & 0.847 & 99.91 & $>$ 99.99 \\
             & 5& 250 & 8 & -5.088 & 1.078 &
            97.36 & 97.40\\
             & 5& 250 & 16 & -6.158 & 0.539 & $>$99.99 & $>$99.99\\
             & 5& 500& 8 & -5.245 & 1.167 &
            97.28 & 98.40\\
            & 5 & 500 & 16 & -6.570 & 0.646 & $>$99.99& $>$99.99 \\
             \hline
            \parbox[t]{2mm}{\multirow{12}{*}{\rotatebox[origin=c]{90}{TPM}}} & 0& 250 & 8 & -2.858 & 0.247 &
            28.30 & 26.00\\
             & 0& 250 & 16 & -2.975& 0.221 & 45.51 & 43.20\\
             & 0& 500& 8 & -2.973 & 0.228 &
            45.29 & 42.10\\
             & 0& 500 & 16 & -3.061 & 0.284 & 58.36& 57.20 \\
             & 2& 250 & 8 & -3.590 & 0.610 &
            83.31 & 82.60\\
             & 2& 250 & 16 & -5.180 & 0.819 & 99.61 & 99.46\\
             & 2& 500& 8 & -3.600 & 0.393 &
            93.65 & 95.92\\
             & 2& 500 & 16 & -5.390 & 0.869 & 99.73& 99.70 \\
             & 5& 250 & 8 & -5.087 & 1.075&
            97.41 & 97.40\\
             & 5& 250 & 16 & -6.150 & 0.539 & $>$99.99 & $>$99.99\\
            & 5 & 500& 8 & -5.257 & 1.138 &
            97.63 & 98.40\\
             & 5& 500 & 16 & -6.569 & 0.646 & $>$99.99& $>$99.99 \\
            \hline
    \end{tabular}
\end{table}

\begin{figure}[t!]
    \centering
     \includegraphics[width = 0.9\columnwidth]{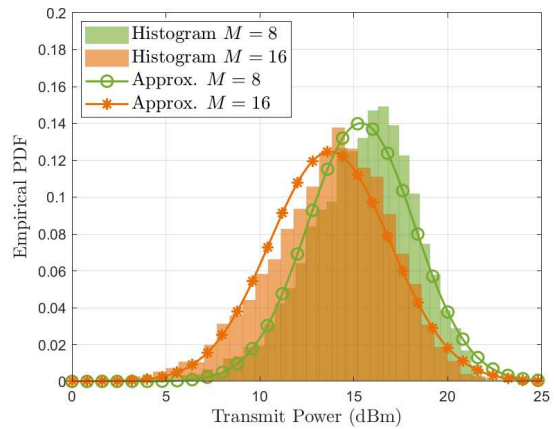}
    \caption{Empirical PDF of the total transmit power and approximation to a Gaussian distribution for 5000 network realizations. All users are subject to Rayleigh fading ($\kappa_k = 0 \ \forall k$). We set $\gamma_{k}^{tar} = 10$ dB $\forall k \ne 0$, and $\xi = 10^{-3}$.}
    \label{fig:9}
\end{figure}

 Fig. \ref{fig:9} shows the (approximately Gaussian) empirical PDF of the total transmit power in dBm when using $M = 8$ and $M = 16$ for $L = 500$ with ZF under Rayleigh fading. The mean transmit power is approximately 16.41 dBm and  14.52 dBm for $M = 8$ and $M = 16$, respectively. This power reduction of about 1.9 dB is the cause that the mean outage probabilities illustrated in Fig. \ref{fig:10} did not experience a larger reduction as previously discussed. Finally, we would like to highlight that under a similar setup with $M = 8$, TPM exhibits an even poorer performance since it requires on average 22.14 dBm of transmit power to satisfy the reliability requirements. Nevertheless, this behavior changes in scenarios with enhanced channel conditions, \textit{e.g.,} Rician fading with $\kappa_0 = 10$, since 12.31 dBm and 11.37 dBm of transmit power are required by ZF and TPM, respectively.

\section{Conclusions}\label{section_5}
In this paper, we considered the I-CSI of multiple eMBB links and the channel measurement’s history of one URLLC user for DL multi-antenna beamforming design. In our proposal, we leveraged the Chernoff bound to stochastically model, impose, and guarantee the reliability requirements of the URLLC user based on its channel history.
Moreover, our proposed precoding design relies on properly modified I-CSI-based precoding methods. We illustrated our approach by adopting ZF and TPM precodings with per-user SINR constraints, whose performance was assessed through simulations. We showed that ZF outperforms TPM in scenarios with poor channel conditions, while TPM exhibits a better performance as the channel becomes more deterministic, \textit{i.e.,} with greater LOS. For instance, in Rayleigh fading with 500 past URLLC measurements, eight antennas at the BS, and for an outage probability target of $10^{-3}$, the mean transmit power of ZF and TPM are 16.41 dBm and 22.14 dBm, respectively. However, in Rician fading with a LOS of 10 dB, the figures drop to 12.31 dBm and 11.37 dBm, respectively. Finally, we determined the confidence levels required to achieve the target outage probabilities, which can be larger than 99\% when operating in favorable LOS conditions.

\bibliographystyle{IEEEtran}
\bibliography{IEEEabrv,References}
\end{document}